\documentclass[10pt,journal,compsoc]{IEEEtran}

\ifCLASSOPTIONcompsoc
  \usepackage[nocompress]{cite}
\else
  \usepackage{cite}
\fi
\ifCLASSINFOpdf
\else
\fi

\newlength{\Oldarrayrulewidth}
\newcommand{\Cline}[2]{%
  \noalign{\global\setlength{\Oldarrayrulewidth}{\arrayrulewidth}}%
  \noalign{\global\setlength{\arrayrulewidth}{#1}}\cline{#2}%
  \noalign{\global\setlength{\arrayrulewidth}{\Oldarrayrulewidth}}}

\usepackage[inline]{enumitem}
\usepackage{csvsimple}
\usepackage{multirow}
\usepackage{subcaption}
\usepackage{csvsimple}
\usepackage{float}
\usepackage[inline]{enumitem}
\usepackage{hang}
\usepackage{amsmath}
\usepackage[pdftex]{graphicx}

\begin{document}
%
\title{A Machine Learning Pipeline Stage for Adaptive Frequency Adjustment}
%
%
%
%

\author{Arash Fouman Ajirlou,~\IEEEmembership{Student Member,~IEEE,}
        Inna~Partin-Vaisband,~\IEEEmembership{Member,IEEE}
\IEEEcompsocitemizethanks{\IEEEcompsocthanksitem A. Fouman was with the Department
of Electrical and Computer Engineering, University of Illinois at Chicago, Chicago,
IL, 60607.\protect\\
E-mail: afouma2@uic.edu
\IEEEcompsocthanksitem I. Parin-Vaisband was with the Department
of Electrical and Computer Engineering, University of Illinois at Chicago, Chicago,
IL, 60607.\protect\\
E-mail: vaisband@uic.edu}
}
\IEEEtitleabstractindextext{%
\begin{abstract}

A machine learning (ML) design framework is proposed for adaptively adjusting clock frequency based on propagation delay of individual instructions. A random forest model is trained to classify propagation delays in real time, utilizing current operation type, current operands, and computation history as ML features. The trained model is implemented in Verilog as an additional pipeline stage within a baseline processor. The modified system is experimentally tested at the gate level in 45 nm CMOS technology, exhibiting a speedup of 70\% and energy reduction of 30\% with coarse-grained ML classification. A speedup of 89\% is demonstrated with finer granularities with 15.5\% reduction in energy consumption.

\end{abstract}

}

\maketitle

\IEEEdisplaynontitleabstractindextext

%
\IEEEpeerreviewmaketitle

\ifCLASSOPTIONcompsoc
\IEEEraisesectionheading{\section{Introduction}\label{Introduction}}
\else
\section{Introduction}
\label{Introduction}
\fi

%
%
%
%
\IEEEPARstart The primary design goal in computer architecture is to maximize the performance of a system under power, area, temperature, and other application-specific constraints. Heterogeneous nature of VLSI systems and the adverse effect of  process, voltage, and temperature (PVT) variations have raised challenges in meeting timing constraints in modern integrated circuits (ICs). To address these challenges, timing guardbands have constantly been increased, limiting the operational frequency of synchronous digital circuits. On the other hand, the increasing variety of functions in modern processors increases delay imbalance among different signal propagation paths. Bounded by critical path delay, these systems are traditionally designed with pessimistically slow clock period, yielding underutilized IC performance. Moreover, power efficiency of these underutilized systems also degrades due to the increasing power leakage. Alternatively, when designed with relaxed timing constraints, integrated systems are prone to functional failures. To simultaneously maintain correct functionality and increase system performance, numerous optimization techniques as well as offline and online models have recently been proposed including: pipelining, multicore computing, dynamic frequency and voltage scaling (DVFS), and ML driven models \cite{Fields02,Zyuban04, kumar03, rahimi17_slot, Hashemi16, Moghadas18,gepner06,hu04, wu04}.

Propagation delay in a processor is a strong function of the type, input operands, and output of the current operation, and  computation history \cite{rahimi17_slot}. Computation history accounts for data overwrite and crosstalk noises. 
  Intuitively, majority of operations are completed within a small portion of the clock period, as determined by the slowest path in the circuit. Based on path delay distribution, as reported in \cite{Hashemi16}, the operational frequency can be doubled for majority ({\it e.g.,} 86.7\% in \cite{Hashemi16}) of instructions in a typical program.

While multicore approaches have been proposed to enhance system performance, the scalability of modern multicore systems is limited by the design complexity of instruction level parallelism and thermal design power constraints \cite{Wang19, Rapp19}.
Thus, speeding a single thread execution is an important cornerstone for enhancing performance in modern ICs \cite{Isci06}. This is, therefore, the primary focus of the proposed approach. To the best of the authors' knowledge, this paper is the first to employ ML for adaptively adjusting the clock frequency at the instruction-level. Note that with the proposed method, the clock frequency is adaptively adjusted {\it per instruction} in real time, yielding a fundamentally different approach as compared with the traditional, task-based dynamic frequency scaling.
The main contributions of this work are as follows:

\begin{enumerate}
\item A systematic flow is proposed and implemented as a unified platform for extracting ML input features from an instruction and classifying the instruction execution delay in real time. 
\item A random forest (RF) model is trained to classify individual instructions into delay classes based on their type, input operands, and the computation history of the system.
\item A new pipeline stage is integrated within a pipelined MIPS processor.
\item The proposed method is synthesized and verified on LegUp \cite{legup} benchmark suite of programs with Synopsys Design Compiler  in 45 nm CMOS technology node.
\end{enumerate}

The rest of the paper is organized as follows. Section \ref{Related Works} describes prior and related work. Section \ref{The proposed ML based frequency adjustment} explains the proposed unified platform and the design methodology. ML algorithms for classification of instruction delay are described in Section \ref{Machine learning models}. In Section \ref{Implementation} the implementation details of the system are introduced. Experimental results are presented in Section \ref{Experimental Results}. Conclusions and future work are discussed in Section \ref{Discussion}, and the paper is summarized in Section \ref{Conclusion}.

\section{Prior and Related Work} \label{Related Works}

Multiple approaches have been proposed for efficiently tuning the operating point ({\it i.e.,} voltage supply and clock frequency) of a system at various levels of a computing system, including application- and task-based methods and instruction-level speculations. 

Predicting timing violations in a constraint-relaxed system is impractical with deterministic approaches, due to the wide dynamic range of input and output signals (typically 32 or 64 bits), variety of operations in a modern processor, and delay dependence on the runtime and physical characteristics of the system ({\it e.g.,} crosstalk noise). ML based approaches for predicting timing violations of individual instructions have recently been proposed,  which consider the impact of input operands and  computation history  on timing violations \cite{rahimi17_slot, rahimi17_clim, FATE18}. While significant for the design process of next generation scalable high performance systems, these approaches have several limitations:
\begin{enumerate} [label={\arabic*}), 
       itemjoin={{\\}}]
       
\item Instruction output is considered as a ML feature and exploited in these systems for predicting the timing characteristics of the individual instructions. These predictions are, however, carried out before the instruction execution, when the instruction output is not yet available, limiting the effectiveness of these methods in practical systems. 
\item The modules under the test are studied separately and evaluated in an isolated test environment without the effects of other processing elements ({\it e.g.,} arithmetic modules, buffers or multiplexers). The high reported accuracy is, therefore, expected to degrade if the methods are applied to a complex system ({\it e.g.,} a practical execution unit).

\item Power and timing overheads due to additional hardware are not considered in these papers.
\end{enumerate}

Granularity of prediction is another primary concern. A bit-level ML based method has been proposed in \cite{rahimi15} for predicting  timing violations with reduced  timing guardbands. While up to 95\% prediction accuracy has been reported with this method, the excessively high, per bit granularity of the ML predictions is expected to exhibit substantial power, area, and timing overheads. These overheads are, however, not evaluated in \cite{rahimi15}.
 Furthermore, a procedure for recovery upon a timing error is not provided and the recovery overheads are also not considered.

As an alternative to fine-grain high-overhead ML methods, multiple coarse-grain schemes for timing error detection and recovery have been proposed to mitigate the adverse effect of the pessimistic design constraints.
A better-than-worst-case design approach has been introduced in \cite{Hashemi16}. With this approach, the clock period is set to a statistically nominal value (rather than worst-case propagation delay) and the history of timing erroneous program counters is kept in a ternary content-addressable memory (TCAM). The TCAM is exploited for predicting timing violations of the instructions based on previous observations. Note that the system only warns against  those timing violations that have been previously recorded. Alternatively, unseen violations are not predicted with this approach.  Owning to the apparent simplicity of this approach, only bi-state operating conditions ({\it i.e.,} nominal and worst-case clock frequencies) can be efficiently utilized with this method. Alternatively, the design complexity and system overheads are expected to significantly increase with the increasing number of frequency domains.

In BandiTS \cite{BandiTS17}, a reinforcement learning approach has been proposed to estimate the timing error probability (TEP) within a program time interval, given timing speculation (TS) ratios, $TSR = {t_{clk}}/{t_{nom}}$ for various values of the reduced clock period $t_{clk}$, and the worst-case clock period $t_{nom}$. The TS-based TEP problem is modeled in \cite{BandiTS17} as the classical multi-armed bandit problem \cite{MAB}, where the TS ratios and TEPs correspond to, respectively, the arms and stochastic rewards. The primary limitation of that work is the lack of details about the hardware implementation and overheads. In addition, the maximum achievable performance gain of only 25\%  has been reported. Furthermore, BandiTS approach exhibits per-task clock granularity and scales the clock frequency for a batch of instructions. Higher performance gain is possible with fine-grain, per instruction clock frequency adjustment, as shown in this paper.

A thermal-aware voltage scaling  has been proposed in \cite{salamat19}. Voltage selection algorithm has been developed and integrated within FPGA synthesis process to aggressively scale the core and block RAM voltages, utilizing the available thermal headroom of the FPGA-mapped design. As a result, 36\% reduction in power consumption has been demonstrated. Driven by workload and thermal power dissipation, this method, however, supports only coarse-grain voltage and frequency scaling.

Predicting program error rate in timing-speculative processors  has been proposed in \cite{assare19}. A statistical model is developed for predicting dynamic timing slack (DTS) at various pipeline stages. The predicted DTS values are exploited to estimate the timing error rate in a program. The implementation overheads, and the potential performance or power consumption gains are, however, not reported with this approach.

An offline model for TS processors has been introduced in \cite{Kruijf10}. This probabilistic model is trained to optimally select a better-than-worst-case, nominal clock frequency. The provided hardware-based speculation, however, does not consider the overall workload or specific finer units, limiting the fidelity of  the method. Alternatively, the adverse effect of process variations on the propagation delay is considered, strengthening the approach in \cite{Kruijf10}. Note that PVT variations are also considered with the proposed approach of classifying instructions into delay intervals in real time, as described in the following sections.

Finally, ML based methods  for modeling system behavior have also been proposed. For example, in \cite{Moghadas18},  linear regression has been leveraged for modeling the aging behavior of an embedded processor based on current instruction and its operands, as well as the computation history and overall circuit switching activity. As a result, the timing guardband designed to compensate for aging in digital circuits can be effectively reduced, in presence of graceful degradation \cite{Moghadas18}. Reallocation of delay budget has, however, not been considered with this method.

ML ICs can exhibit a prohibitively high power consumption and physical size. Furthermore, ML ICs can introduce additional delay and increase design complexity, depending upon the application characteristics. To efficiently exploit ML methods for managing frequency in modern processors, delay, power, and area of ML ICs should be considered.

\begin{figure}[b]
  \includegraphics[width=\linewidth]{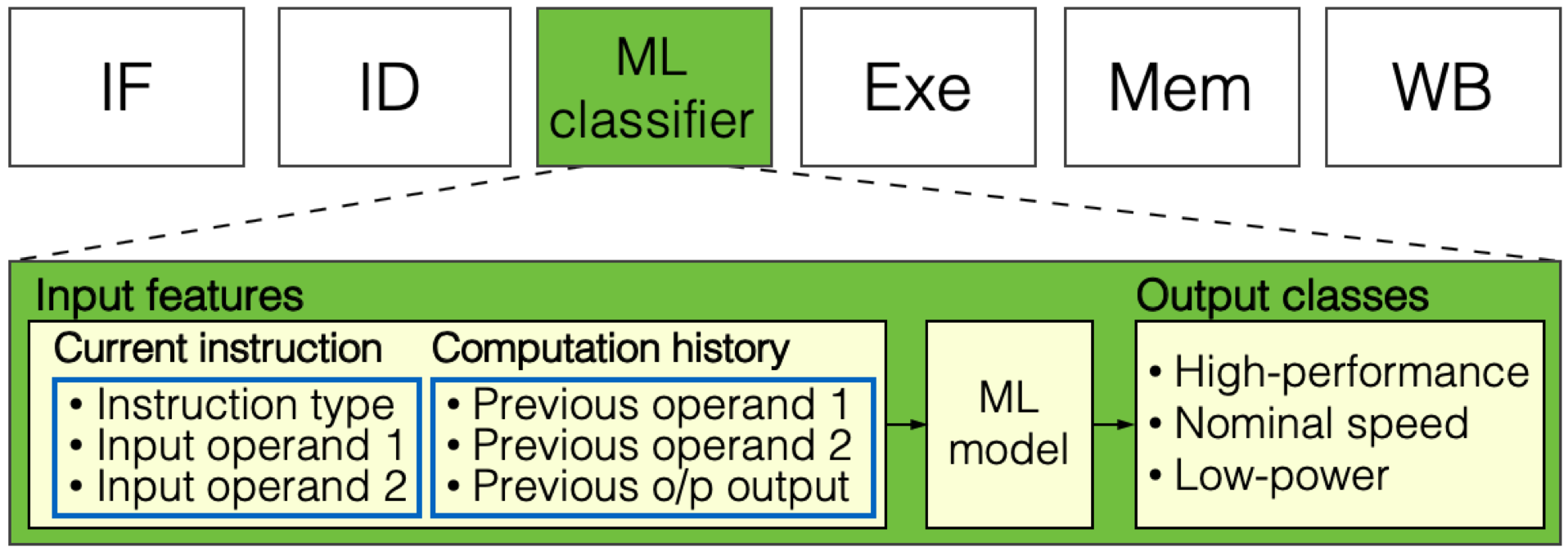}
  \caption{The proposed pipeline with the additional ML stage. In this configuration, six ML features and three delay classes are illustrated.}
  \label{fig:pipeline}
\end{figure}

\section{The proposed ML based frequency adjustment} \label{The proposed ML based frequency adjustment}

In this paper, a design methodology is proposed for ML driven adjustment of operational frequency in pipeline processors. With the proposed method, individual instructions are classified into the corresponding propagation delay classes in real time, and  the clock frequency is accordingly adjusted to reduce the gap between the actual propagation delay and the clock period. The classes are defined by segmenting the worst-case clock period into shorter delay fragments. Each class is characterized by a specific supply voltage and clock frequency. The primary design objective is to maximize system performance within an allocated energy budget. The overall delay and energy consumption are evaluated \underline{with} the additional ML components, and both the correct and incorrect predictions. The proposed scalable framework allows for other control configurations to be defined in a similar manner for different design objectives. The real-time clock adjustment is enabled by the recent advancement in clock management circuits \cite{Jia19_ISSCC}. 

In order to evaluate this method, a pipelined, 32-bit MIPS processor (TigerMIPS\cite{TigerMIPS}) is utilized as the baseline processor. The ML classifier is designed as an additional pipeline stage within the pipelined MIPS processor, as shown in Fig. \ref{fig:pipeline}. The inputs to the additional ML pipeline stage are the current instruction and its operands, as well as the computation history, as defined by the toggled inputs bits ({\it i.e.,} current inputs are XORed with the previous inputs) and output of the previous operation. The choice of these parameters is in accordance with the results in \cite{rahimi17_slot} and \cite{Moghadas18}. These inputs are utilized as ML features for predicting the delay class of the current instruction based on the trained ML model.  It is important to note that more complex, slower  ML models can also be trained with this methodology, as long as the design complexity and hardware costs of the final system meet the specified constraints. To meet  the overall system throughput constraints, the trained models can be implemented as multiple pipeline stages, mitigating the additional latency  introduced by the ML functions. Finally, the granularity of the output delay ({\it e.g.,} three delay classes are illustrated in Fig. \ref{fig:pipeline}) can be varied to meet the timing constraints within the energy budget.

\begin{figure*}[h]
  \includegraphics[width=\linewidth]{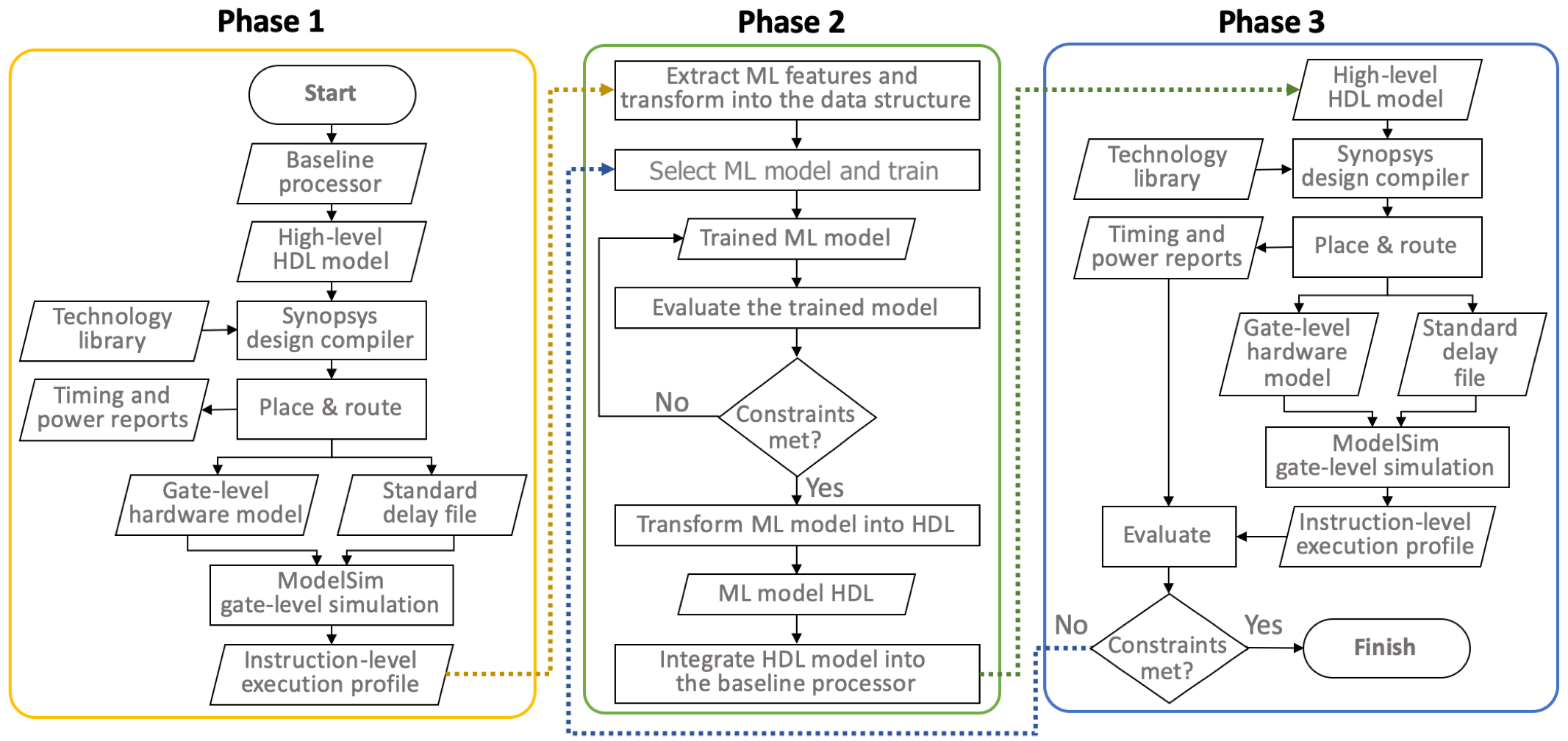}
  \caption{Systematic flow for designing ML predictor within a typical pipelined processor.}
  \label{fig:sysFlow}
\end{figure*}

A systematic flow has been developed, implemented, and verified on TigerMIPS with LegUp benchmark suite. The flow comprises three primary phases, as shown in Fig. \ref{fig:sysFlow}. The individual phases are described in the following subsections.

\subsection{Phase 1: Baseline processor synthesis and profiling }  \label{Phase 1: Baseline processor synthesis and profiling } 
First, the high-level hardware description language (HDL) model of the baseline processor is synthesized into gate-level description model. During this phase, timing information is generated in the IEEE standard delay format (SDF). Based on this information, the gate-level simulation (GLS) is performed and the instruction-level execution profile is generated. A profile comprises a list of instructions, the fetched or forwarded operands, the output of the operations, and the propagation delays. In addition to the execution profile, post place-and-route (PAR) reports, including timing and power information, are collected in this phase.

\begin{figure}[h]
	\centering
	\begin{subfigure}{\columnwidth} 
		\includegraphics[width=\columnwidth]{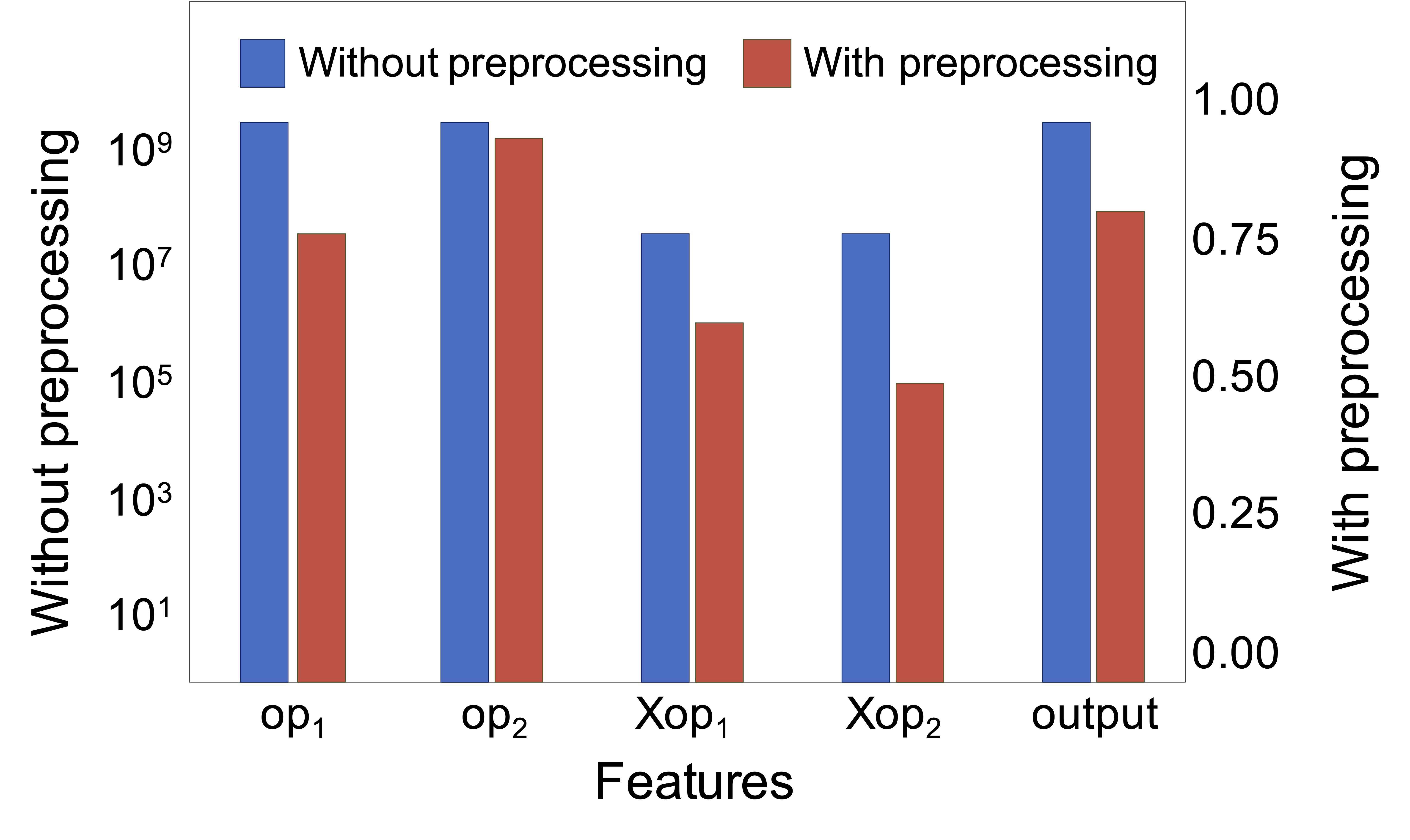}
		\caption{} 
	\end{subfigure}
	\vspace{0.1mm} 
	\begin{subfigure}{\columnwidth} 
		\includegraphics[width=\columnwidth]{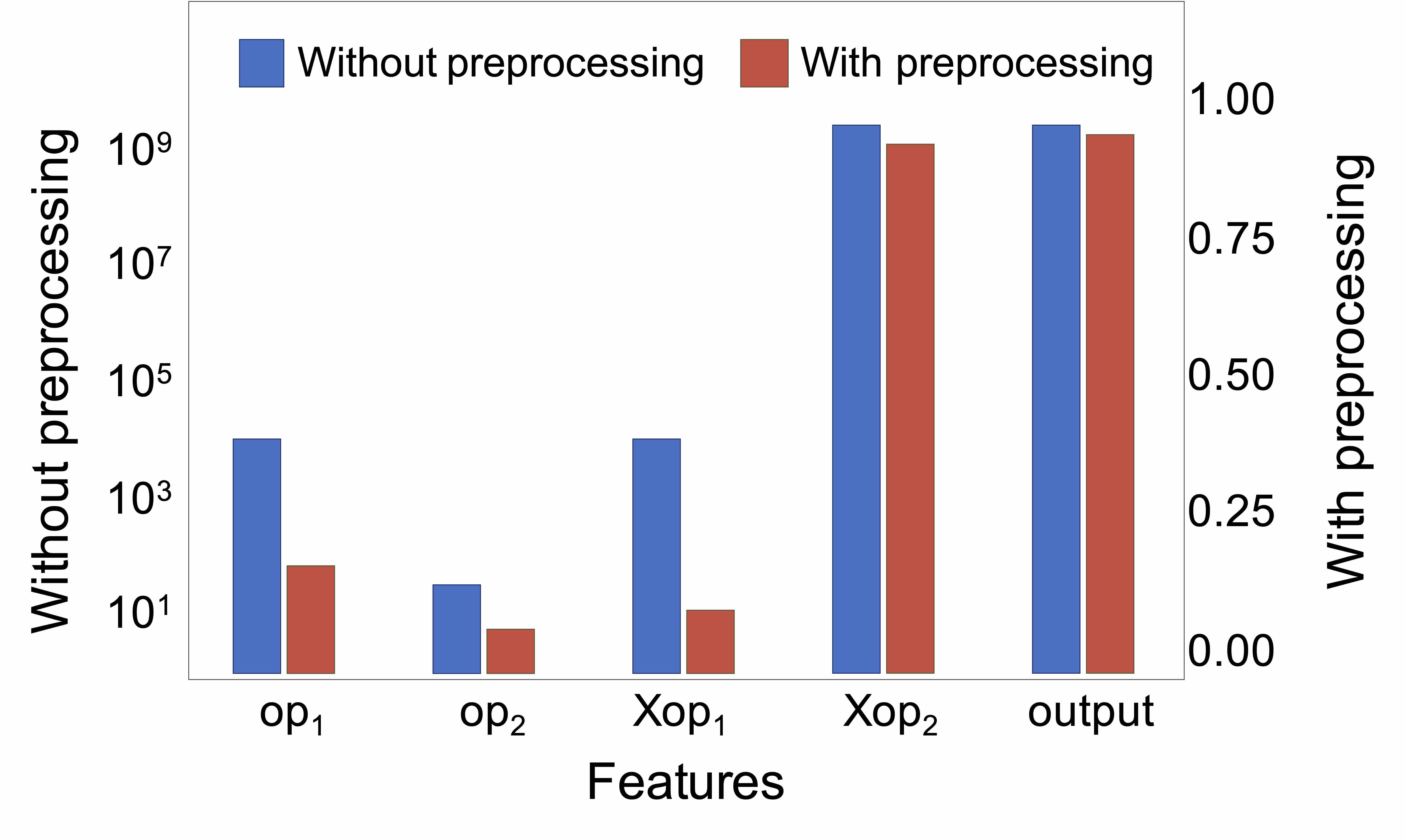}
		\caption{} 
	\end{subfigure}
	  \caption{A typical feature vector with and without the ML preprocessing, (a) for arithmetic operation with immediate operand, and (b) for logical operation. Note that the values without preprocessing are shown on a logarithmic scale, while the values with preprocessing are shown on a linear scale.}
\label{fig:features}
\end{figure}

\subsection{Phase 2: ML training} \label{Phase 2: ML training}
In this phase, the gate-level profiles from Phase 1 are parsed and utilized as ML features. Based on the extracted features, a preferred ML model is trained in Python with Scikit-learn ML library\cite{scikit-learn}. A HDL code ({\it e.g.,} Verilog in this paper) of the trained model is generated and integrated within the baseline processor as a single (or multiple) pipeline stage(s) between the decode and execute stages (see Fig. \ref{fig:pipeline}).

\subsection{Phase 3: Verification and Evaluation} \label{Phase 3: Verification and Evaluation}
During this phase, the modified high-level HDL model of the system with the ML pipeline stage is synthesized and profiled, as described in Phase 1. To guarantee functional correctness,  the output signal is double-sampled to detect timing violations, and timing-erroneous instructions are re-executed with the worst-case clock frequency. Similar to the baseline iteration, the post PAR reports are extracted for evaluating the timing and energy characteristics of the system. Finally, the profiling of the modified system is executed during this phase to evaluate the overall speedup of the system.

To optimize the final solution in terms of the  operational frequency and energy consumption, the proposed flow is executed iteratively with various ML algorithms and clock fragments, as shown with the feedback in Fig. \ref{fig:sysFlow}. The clock signal of the pipeline registers is assumed to be near-instantly switched based on the individual classification results, as has been experimentally demonstrated in \cite{Jia19_ISSCC}.

\section{Machine learning models} \label{Machine learning models}

Owing to the unique learning characteristics and hardware trade-offs of neural networks (NNs), support vector machines (SVMs), and random forest (RF) models, all these ML models are considered in this paper. Each model is trained based on the instruction profiles extracted from a synthetically generated dataset of 3,000 random instructions per class. The delay boundaries of the individual classes are experimentally determined with respect to the worst-case delay of 4 ns as follows: \{[0.0,2.2],(2.2,4.0]\} for the two-class configuration, \{[0.0,1.8],(1.8,2.6],(2.6,4.0]\} for the three-class configuration, and \{[0.0,1.0],(1.0,2.0],(2.0,3.0],(3.0,4.0]\} for the four-class configuration.

The feature vector of the $i^\text{th}$ instruction comprises six elements, $x_i = (instr, op_1, op_2, Xop_1, Xop_2, output)$.
The first feature, $instr$, comprises four subfeatures, representing the type of the operation in one-hot format,
\[
instr = \left\{\begin{array}{lr}
  1000, & \text{if arithmetic}\\
  0100, & \text{if arithmetic with immediate operand}\\
  0010, & \text{if logical}\\
  0001, & \text{if multiplication or division}
\end{array}
\right.
\]
The subsequent four elements are defined by the operands. The features $op_1$ and $op_2$ are the first and second operands of the instruction, and the features $Xop_1$ and $Xop_2$ are the XORed values of the first and second operands with their respective previous values. The last feature, $output$, is the output of the preceding instruction. The last three elements of the feature vector are exploited to capture the effect of computation history on the instruction delay. Note that the operands and output of the preceding instruction are 32-bit long, as determined by the 32-bit baseline processor utilized in this work. Thus, the distribution of these features significantly differs from the distribution of the operation type subfeatures. To balance the overall distribution of the individual features, the input features are preprocessed and scaled to follow a normal distribution using {\it quantile transformer} in Python scikit-learn library. An example of operand and output features with and without the transformation is shown in Fig. \ref{fig:features} for arithmetic and logical instructions. Note that the type subfeatures remain unchanged.

To evaluate the efficiency and efficacy of the proposed method, propagation delay classification is investigated with three common ML algorithms: NN, SVM, and RF. The configuration of each of the three ML models is described in the following subsections, including the hyperparameters, performance, and hardware costs of the individual ML algorithms. All the algorithms are five-fold cross-validated based on three thousand randomly generated instructions per class. While finding an effective metric for stability of the evaluation is still an open question, k-fold cross-validation with $5 \leq K \leq 20$ is typically used, as these {\it K} values have been demonstrated to simultaneously minimize the bias and variance across many studied test sets \cite{James13, Kuhn13, Kohavi, Forman}. Thus, $K = 5$ is used in this work. ML accuracy is reported as the F1-score of delay classification and the resultant speedup for each benchmark program has been considered in determining the performance of each ML algorithm. Hardware cost is evaluated as the number of additional transistors required for implementing the individual ML algorithms and has also been considered in determining the performance of the ML algorithms. Among the evaluated ML algorithms, the RF classifier is preferred in this work due to the favorable tradeoff between the performance gain and hardware costs, as well as the relative simplicity of the RF algorithm, as explained in the following subsections.

\begin{table}[h]
\caption{Top (within 1\% of the highest F1-score) NN configurations and their respective performance metrics ({\it i.e.,} speedup, hardware cost (in million transistors), and speedup per hardware metric (SPH)). }
\small\addtolength{\tabcolsep}{-5pt}
\resizebox{\columnwidth}{!}{
\begin{tabular}{| | c ||c  | c | c | c | c | c | c ||}
\hline
    \bfseries    & 
    \multicolumn{7}{c||}{ \bfseries 4 classes } \\ 
        \cline{2-8} 
    & Activation & Solver & Neurons & F1-score & Speedup & HW cost & SPH 
    \begin{filecontents*}{gridSearch_nn2.csv}
a,b,c,d,e,f,g,h,i,j,k,l,m,n,o,p,q,r,s,t,u,v
1,tanh,lbfgs,10,0.859,1.915,2.834,0.676,logistic,lbfgs,$20\times15$,0.923,1.645,11.461,0.143,logistic,adam,$20\times15$,0.972,1.682,11.134,0.151
2,relu,adam,$20\times5$,0.858,1.940,6.543,0.297,tanh,adam,$20\times15$,0.922,1.642,11.461,0.143,identity,lbfgs,$20\times10$,0.972,1.682,8.730,0.193
3,relu,adam,$20\times10$,0.856,1.882,9.166,0.205,logistic,lbfgs,$20\times10$,0.921,1.643,8.948,0.184,identity,lbfgs,20,0.972,1.682,4.797,0.351
4,tanh,adam,$20\times10$,0.856,1.911,9.166,0.208,relu,lbfgs,$20 \times 5$,0.920,1.642,6.434,0.255,identity,lbfgs,$20 \times 5$,0.972,1.682,6.326,0.266
5,tanh,adam,$20\times15$,0.853,1.916,11.789,0.163,logistic,lbfgs,15,0.920,1.642,3.925,0.418,identity,lbfgs,10,0.972,1.682,2.398,0.701
6,tanh,lbfgs,20,0.850,1.934,5.671,0.341,relu,adam,20,0.919,1.643,5.234,0.314,identity,lbfgs,5,0.972,1.682,1.198,1.404
7,relu,lbfgs,20,0.848,1.940,5.671,0.342,tanh,adam,$20 \times 5$,0.919,1.642,6.434,0.255,identity,lbfgs,$20\times15$,0.972,1.682,11.134,0.151
8,logistic,lbfgs,20,0.848,1.836,5.671,0.324,tanh,lbfgs,5,0.918,1.643,1.307,1.257,relu,adam,10,0.972,1.682,2.398,0.701
9,tanh,adam,$20 \times 5$,0.845,1.946,6.543,0.297,relu,adam,$20 \times 5$,0.914,1.643,6.434,0.255,identity,lbfgs,15,0.972,1.681,3.598,0.467
10,relu,adam,$20\times15$,0.844,1.951,11.789,0.165,tanh,adam,$20\times10$,0.914,1.643,8.948,0.184,identity,adam,5,0.969,1.678,1.198,1.400
\end{filecontents*}
\csvreader[head to column names]{gridSearch_nn2.csv}{}
    {\\\hline \a &  \b & \c & \d & \e & \f & \g & \h }
        \\   \Cline{0.65pt}{1-8}
    \multicolumn{4}{||c|}{Average	}							&  0.852	&	1.917	&	7.484	&	0.302   \\
   \cline{1-8}
   \multicolumn{4}{||c|}{Positive standard  deviation($\sigma^+$) }	&   0.002	&	0.012	&	1.634	&	0.095   \\
   \cline{1-8}
   \multicolumn{4}{||c|}{Negative standard  deviation($\sigma^-$) }	&   0.002	&	0.018	&	0.961	&	0.040  
 \\\hline   \hline

   \bfseries    & 
   \multicolumn{7}{c||}{ \bfseries 3 classes } \\ 
        \cline{2-8} 
    & Activation & Solver & Neurons & F1-score & Speedup & HW cost & SPH 
    \csvreader[head to column names]{gridSearch_nn2.csv}{}
    {\\\hline \a &  \i & \j & \k & \l & \m & \n & \o }
        \\   \Cline{0.65pt}{1-8}
   \multicolumn{4}{||c|}{Average	}							&  0.919	&	1.643	&	7.059	&	0.341   \\
   \cline{1-8}
   \multicolumn{4}{||c|}{Positive standard  deviation($\sigma^+$) }	&   0.001	&	3.7E-4	&	1.694	&	0.460   \\
   \cline{1-8}
   \multicolumn{4}{||c|}{Negative standard  deviation($\sigma^-$) }	&   0.001	&	4.0E-4	&	1.148	&	0.049  
    \\\hline   \hline

   \bfseries    & 
   \multicolumn{7}{c||}{ \bfseries 2 classes } \\ 
        \cline{2-8} 
    & Activation & Solver & Neurons & F1-score & Speedup & HW cost & SPH 
    \csvreader[head to column names]{gridSearch_nn2.csv}{}
    {\\\hline \a &  \p & \q & \r & \s & \t & \u & \v }
        \\   \Cline{0.65pt}{1-8}
   \multicolumn{4}{||c|}{Average	}							&  0.972	&	1.681	&	5.291	&	0.579   \\
   \cline{1-8}
   \multicolumn{4}{||c|}{Positive standard  deviation($\sigma^+$) }	&   1.0E-4	&	1.8E-4	&	2.252	&	0.294   \\
   \cline{1-8}
   \multicolumn{4}{||c|}{Negative standard  deviation($\sigma^-$) }	&   0.003	&	0.002	&	1.217	&	0.137 
    \\\hline 

\end{tabular}
}
\label{table:NN}
\end{table}

\subsection{Neural Networks} \label{Neural Networks}

NNs excel in learning complex hidden patterns in large datasets and have exhibited a particular supremacy in vision and text applications as compared with classical ML algorithms. Following this success, promising results have been shown with NNs in various hardware related applications \cite{NN1,NN2,NN3}. 

To determine the preferred set of hyperparameters for the two-, three-, and four-class NN models, a grid search is executed for each multiclass NN over the following ranges: 
\begin{enumerate}
\item Identity, tanh, logistic, and ReLu activation functions,
\item Stochastic gradient descent \cite{gradientDescent}, lbfgs (a limited memory BFGS quasi-Newton optimization algorithm \cite{bfgs}), and Adam (an adaptive learning rate optimization algorithm \cite{adam}) solvers, and 
\item A single $m$-neuron hidden layer ($m \in \{ 5, 10, 15, 20\}$) and two hidden NN layers with $m_1$ and $m_2$ neurons in, respectively, the first and second layers ($m_1 \times m_2 \in \{20 \times 5, 20 \times 10, 20 \times 15 \}$). 
\end{enumerate}
The networks are trained using backpropagation algorithm for 200 epochs until convergence with quasi-newton optimizer. Note that the number of neurons in the input and output layers is determined by, respectively,  the number of ML features  (nine, including the four instruction type subfeatures) and the number  of ML classes (two, three, and four). The top ten grid search results (within 1\% of the highest F1-score) are listed in Table \ref{table:NN} for each of the multiclass NNs in the descending order of the F1-scores. The hardware cost is determined based on the number of transistors comprising the NN adders and multipliers. The transistor count for the individual NN adders and multipliers is determined based on \cite{transistorCount}. The number of multipliers, $N_{MULT}$, and adders, $N_{ADD}$, in a NN with $L$ layers is determined, respectively, as,
\begin{equation}
N_{MULT} = \sum_{i=1}^{L}m_i \cdot v_i, \\
\end{equation}
and\\
\begin{equation}
N_{ADD} = \sum_{i=1}^{L}m_i \cdot (v_i-1),
\end {equation}
where $m_i$ is the number of neurons in each layer, and $v_i$ is the size of the input vector to each layer (or the feature vector size in the input layer).

The speedup per hardware cost (SPH) is also listed in Table \ref{table:NN} for each of the NN configurations. These top NN results are compared with the SVM and RF top results, as described at the end of this section.  As a general rule, learning capacity of a NN increases with the network complexity ({\it i.e.,} number of neurons and number of layers). For a NN to be competitive with or outperform a classical ML algorithm, a large number of neurons and layers is required, significantly increasing the system complexity and hardware overhead of the NN based solutions.

\subsection{Support Vector Machines} \label{Support Vector Machines}

SVM classifier generates an optimal hyperplane which separates data samples in feature space with the objective to minimize the classification error. Linear SVM can only classify a linearly-separable data. Alternatively, to learn complex nonlinear data patterns, SVM can be combined with a {\it kernel trick}, enabling the feature transformation into linearly separable space \cite{kernelTrick}. In this work, a grid search is performed over the following kernel SVM hyperparameters: 
\begin{enumerate}
\item Linear, polynomial, and radial basis function (rbf) kernels, 
\item Integer degree of flexibility of the polynomial decision boundary, $d \in [2,15]$, and 
\item The influence on the model of a single sample in a training set with $N$ features and variance $Var$ by scaling ({\it i.e.,} $gamma = 1/(N\cdot Var)$) or not scaling ({\it i.e.,} $gamma = 1/N$) the kernel coefficient, $gamma$.
\end{enumerate}
 The sets of hyperparameters with the highest F1-scores are listed in Table \ref{table:SVM}. The speedup, hardware cost, and SPH metric are also listed in the table for all the SVM configurations.  SVM hardware cost is determined as the number of transistors, based on the method presented in \cite{svmCost}. SVM often exhibits excellent performance as compared with other learning algorithms at the expense of higher computational and design complexity, and accordingly higher power and area overheads \cite{mohsenin16}. These tradeoffs are discussed at the end of this section.

\begin{table}[t]
\caption{Top (within 1\% of the highest F1-score) SVM configurations and their respective performance metrics ({\it i.e.,} speedup, hardware cost (in million transistors), and speedup per hardware metric (SPH)). }

\small\addtolength{\tabcolsep}{-4pt}
\resizebox{\columnwidth}{!}{
\begin{tabular}{| | c ||c  | c | c | c | c | c | c ||}
\hline
    \bfseries    & 
    \multicolumn{7}{c||}{ \bfseries 4 classes } \\ 
        \cline{2-8} 
       & kernel & degree & gamma & F1-score &Speedup & HW cost & SPH  
    \begin{filecontents*}{gridSearch_svm2.csv}
a,b,c,d,e,f,g,h,i,j,k,l,m,n,o,p,q,r,s,t,u,v
1,poly,5,scale,0.837,1.873,1323.343,0.001,poly,5,scale,0.876,1.604,755.765,0.002,rbf,N/A,scale,0.957,1.680,28.616,0.059
2,poly,4,scale,0.834,1.899,1307.230,0.001,poly,4,scale,0.875,1.617,715.028,0.002,poly,4,scale,0.956,1.679,169.482,0.010
3,poly,6,scale,0.833,1.889,1352.670,0.001,poly,3,scale,0.875,1.617,675.647,0.002,poly,5,scale,0.955,1.681,181.843,0.009
4,poly,3,scale,0.828,1.910,1291.739,0.001,rbf,N/A,scale,0.875,1.617,119.954,0.013,poly,6,scale,0.954,1.683,198.685,0.008
5,poly,7,scale,0.827,1.923,1412.040,0.001,poly,7,scale,0.873,1.627,847.086,0.002,poly,3,auto,0.954,1.674,323.000,0.005
6,poly,8,scale,0.826,1.915,1476.320,0.001,poly,2,scale,0.872,1.618,677.269,0.002,poly,3,scale,0.953,1.677,158.120,0.011
7,poly,9,scale,0.823,1.908,1534.991,0.001,poly,6,scale,0.872,1.601,800.553,0.002,poly,8,scale,0.952,1.683,225.245,0.007
8,rbf,N/A,scale,0.822,1.916,228.524,0.008,poly,8,scale,0.870,1.613,900.614,0.002,poly,7,scale,0.952,1.678,212.632,0.008
9,poly,10,scale,0.819,1.814,1596.913,0.001,rbf,N/A,auto,0.869,1.617,135.990,0.012,poly,2,scale,0.951,1.666,145.830,0.011
10,poly,11,scale,0.813,1.793,1654.040,0.001,poly,9,scale,0.869,1.660,948.560,0.002,rbf,N/A,auto,0.951,1.629,27.952,0.058
\end{filecontents*}
\csvreader[head to column names]{gridSearch_svm2.csv}{}
    {\\\hline \a &  \b & \c & \d & \e & \f & \g & \h }
        \\   \Cline{0.65pt}{1-8}
   \multicolumn{4}{||c|}{Average	}							&  0.826	&	1.884	&	1317.781	&	0.002   \\
   \cline{1-8}
   \multicolumn{4}{||c|}{Positive standard  deviation($\sigma^+$) }	&   0.003	&	0.010	&	74.701	&	0.006   \\
   \cline{1-8}
   \multicolumn{4}{||c|}{Negative standard  deviation($\sigma^-$) }	&   0.003	&	0.038	&	363.206	&	2.3E-4
 \\\hline   \hline

    \bfseries    & 
    \multicolumn{7}{c||}{ \bfseries 3 classes } \\ 
        \cline{2-8} 
       & kernel & degree & gamma & F1-score &Speedup & HW cost & SPH  
    \csvreader[head to column names]{gridSearch_svm2.csv}{}
    {\\\hline \a &  \i & \j & \k & \l & \m & \n & \o}
        \\   \Cline{0.65pt}{1-8}
   \multicolumn{4}{||c|}{Average	}							&  0.873	&	1.619	&	657.647	&	0.004   \\
   \cline{1-8}
   \multicolumn{4}{||c|}{Positive standard  deviation($\sigma^+$) }	&   0.001	&	0.021	&	57.771	&	0.006   \\

   \cline{1-8}
   \multicolumn{4}{||c|}{Negative standard  deviation($\sigma^-$) }	&   0.001	&	0.003	&	374.579	&	7.4E-4    
 \\\hline   \hline

    \bfseries    & 
    \multicolumn{7}{c||}{ \bfseries 2 classes } \\ 
        \cline{2-8} 
       & kernel & degree & gamma & F1-score &Speedup & HW cost & SPH  
    \csvreader[head to column names]{gridSearch_svm2.csv}{}
    {\\\hline \a & \p & \q & \r & \s & \t & \u &\v  }
        \\   \Cline{0.65pt}{1-8}
   \multicolumn{4}{||c|}{ Average }		  					&  0.954		&	1.673	&	167.141	&	0.019   \\
   \cline{1-8}
   \multicolumn{4}{||c|}{Positive standard  deviation($\sigma^+$) }	&   9.2 E-4	&	0.003	&	29.323	&	0.028   \\
   \cline{1-8}
   \multicolumn{4}{||c|}{Negative standard  deviation($\sigma^-$) }	&   8.3 E-4	&	0.022	&	49.4330	&	0.004
    \\\hline

\end{tabular}
}
\label{table:SVM}
\end{table}

\begin{table}[t]
\caption{Top (within 1\% of the highest F1-score) RF configurations and their respective performance metrics ({\it i.e.,} speedup, hardware cost (in million transistors), and speedup per hardware metric (SPH)). }

\small\addtolength{\tabcolsep}{-5pt}
\resizebox{\columnwidth}{!}{
\begin{tabular}{| | c ||c  | c | c | c | c | c ||}
\hline
    \bfseries    & 
    \multicolumn{6}{c||}{ \bfseries 4 classes } \\ 
        \cline{2-7} 
    & max\_depth & n\_estimator & F1-score & Speedup & HW cost & SPH  
    \begin{filecontents*}{gridSearch_rf2.csv}
a,b,c,d,e,f,g,h,i,j,k,l,m,n,o,p,q,r,s
1,30,50,0.852,1.835,0.177,10.357,20,50,0.949,1.879,0.156,12.040,40,50,0.981,1.688,0.192,8.785
2,10,200,0.850,1.925,0.480,4.012,30,100,0.947,1.856,0.354,5.238,30,200,0.981,1.686,0.709,2.380
3,30,200,0.850,1.889,0.709,2.666,20,200,0.946,1.851,0.624,2.966,30,100,0.981,1.683,0.354,4.750
4,10,100,0.849,1.913,0.240,7.978,40,200,0.945,1.847,0.768,2.403,40,200,0.981,1.686,0.768,2.193
5,50,200,0.849,1.833,0.815,2.249,40,50,0.944,1.843,0.192,9.591,10,50,0.981,1.686,0.120,14.058
6,50,100,0.846,1.856,0.407,4.554,50,200,0.944,1.839,0.815,2.257,20,200,0.981,1.683,0.624,2.696
7,20,200,0.845,1.874,0.624,3.004,10,200,0.944,1.848,0.480,3.853,10,200,0.980,1.686,0.480,3.515
8,50,50,0.843,1.836,0.204,9.011,30,200,0.942,1.814,0.709,2.561,50,200,0.980,1.687,0.815,2.070
9,30,100,0.842,1.838,0.354,5.187,10,100,0.940,1.828,0.240,7.621,10,100,0.980,1.685,0.240,7.024
10,10,50,0.840,1.902,0.120,15.859,10,50,0.939,1.826,0.120,15.224,20,100,0.979,1.683,0.312,5.393
\end{filecontents*}
\csvreader[head to column names]{gridSearch_rf2.csv}{}
    {\\\hline \a &  \b & \c & \d & \e & \f & \g  }
        \\   \Cline{0.65pt}{1-7}
   \multicolumn{3}{||c|}{ Average }		  					&  0.847	&	1.870	&	0.413	& 	6.488  \\
   \cline{1-7}
   \multicolumn{3}{||c|}{Positive standard  deviation($\sigma^+$) }	&   0.002	&	0.016	&	0.137	& 	2.638  \\
   
   \cline{1-7}
   \multicolumn{3}{||c|}{Negative standard  deviation($\sigma^-$) }	&  0.002	&	0.014	&	0.078	&	1.250 
   \\\hline   \hline

    \bfseries    & 
    \multicolumn{6}{c||}{ \bfseries 3 classes } \\ 
        \cline{2-7} 
    & max\_depth & n\_estimator & F1-score & Speedup & HW cost & SPH  
    \csvreader[head to column names]{gridSearch_rf2.csv}{}
    {\\\hline \a & \h & \i & \j & \k & \l & \m  }
        \\   \Cline{0.65pt}{1-7}
   \multicolumn{3}{||c|}{ Average }		  					&  0.944	&	1.843	&	0.446	&	6.375   \\
   \cline{1-7}
   \multicolumn{3}{||c|}{Positive standard  deviation($\sigma^+$) }	&   0.002	&	0.008	&	0.117	&	2.764   \\
   \cline{1-7}
   \multicolumn{3}{||c|}{Negative standard  deviation($\sigma^-$) }	&   0.002	&	0.007	&	0.111	&	1.360 
 \\\hline   \hline

    \bfseries    & 
    \multicolumn{6}{c||}{ \bfseries 2 classes } \\ 
        \cline{2-7} 
    & max\_depth & n\_estimator & F1-score & Speedup & HW cost & SPH  
    \csvreader[head to column names]{gridSearch_rf2.csv}{}
    {\\\hline \a & \n & \o & \p & \q & \r & \s }
        \\   \Cline{0.65pt}{1-7}
   \multicolumn{3}{||c|}{ Average }		  					&  0.980		&	1.685	&	0.461	&	5.286  \\
   \cline{1-7}
   \multicolumn{3}{||c|}{Positive standard  deviation($\sigma^+$) }	&   2.0E-04	&	0.001	&	0.111	&	2.401   \\
   \cline{1-7}
   
   \multicolumn{3}{||c|}{Negative standard  deviation($\sigma^-$) }	&   4.3E-04	&	0.001	&	0.104	&	1.034   

    \\\hline   

\end{tabular}
}
\label{table:RF}
\end{table}

\subsection{Random Forest} \label{Random Forest}

RF classifier is an ensemble of decision tree classifiers. The input samples are split into multiple sample subsets and each decision tree is trained on one training subset. The final classification decision for each sample is made based on the result of averaging the individual tree decisions ({\it i.e.,} ensembling). RF models benefit from the accuracy, training speed, and interpretability of the decision tree model, while the ensembling mitigates the overfitting, otherwise common to decision tree classifier. RF is often preferred in scientific and practical applications \cite{RF, rahimi17_slot}. The computational and hardware complexity of RF is a strong function of the number and depth of the decision trees. The depth of the individual trees is dependent on the number of features and their correlation. In this work, a RF grid search is performed over the following ranges of hyperparameters: \\
\begin{enumerate}
\item Number of trees in the forest, $n\_estimators \in \{ 1, 10, 50, 100, 200\}$,  
\item Maximum number of levels in each tree, $max\_depth \in \{10, 20, 30, 40, 50\}$. 
\end{enumerate}The results of the top estimators (within 1\% of the highest F1-score) are listed in Table \ref{table:RF}. The hardware cost of an RF classifier is evaluated based on the number of required comparators, $\mathcal{O}(n\_estimators \times \log_2(max\_{depth}))$, and reported in terms of the total number of RF transistors. Transistor count for a single comparator is determined based on \cite{comparatorCost}.

\subsection{ML Algorithm Tradeoffs}
\begin{figure}[t]
  \includegraphics[width=\linewidth]{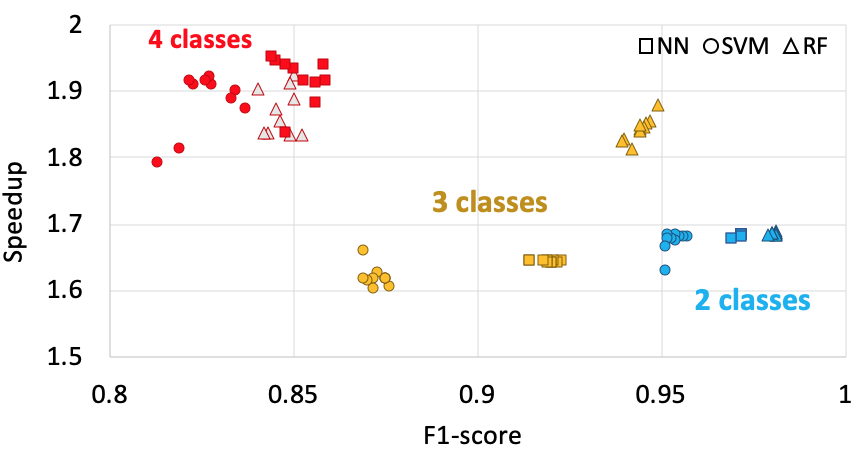}
  \caption{Speedup vs. F1 for two-, three-, and four-class configurations based on Tables \ref{table:NN}-\ref{table:RF}.}
  \label{fig:spdupvsF1}
\end{figure}

\begin{figure}[h]
  \includegraphics[width=8.5cm]{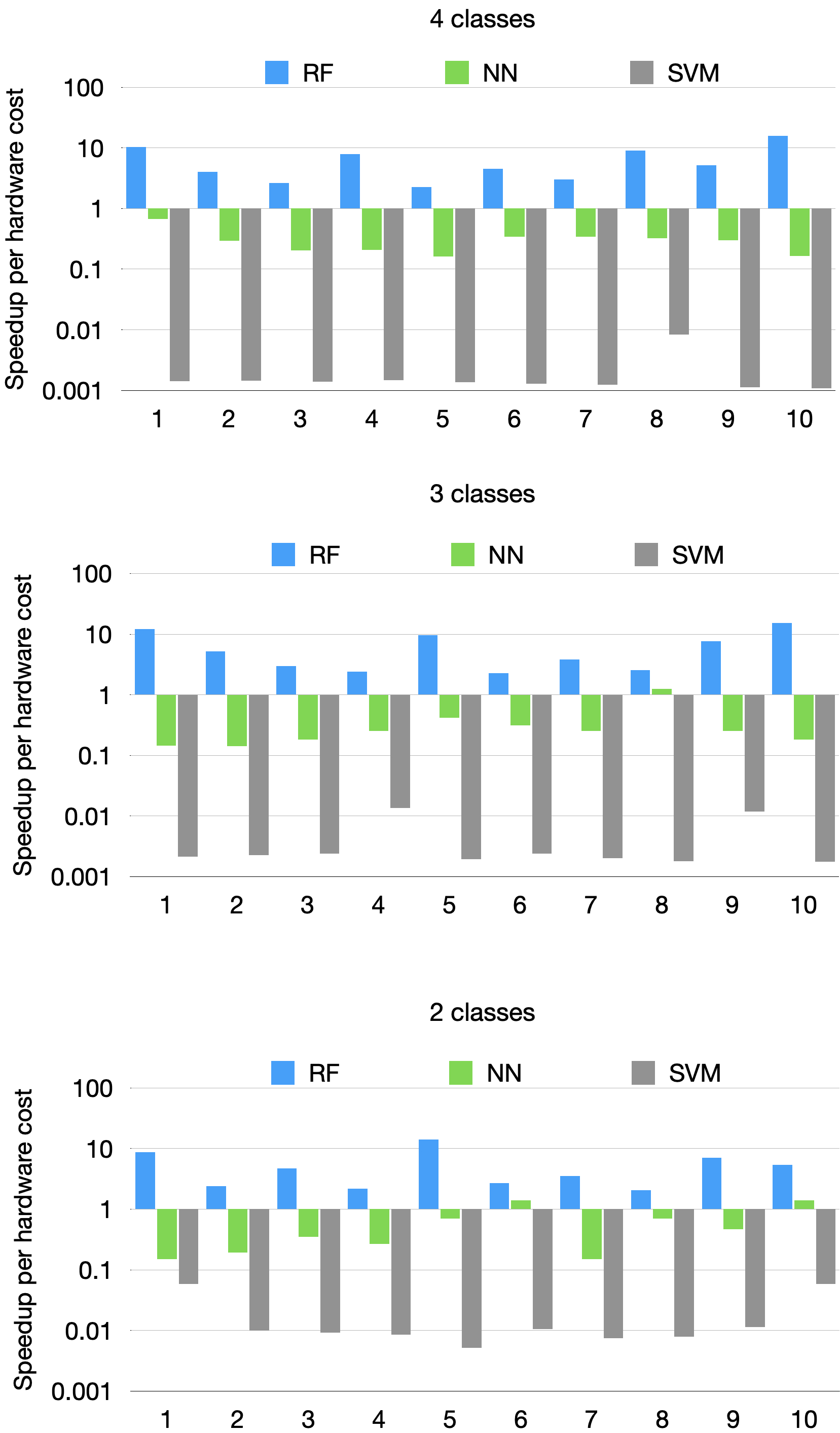}
  \caption{Speedup per hardware cost (SPH) for two-, three, and four-class configurations. The hardware cost is evaluated based on the number of transistors needed to realize each classifier. The SPH performance is highest with RF classifier as compared with the SVM and NN based classifiers for each of the classifier configurations.  Numbers correspond to data listed in Tables \ref{table:NN}, \ref{table:SVM}, and \ref{table:RF}. }
  \label{fig:hwc_spd}
\end{figure}

The tradeoffs between the speedup and F1-score are summarized in Fig. \ref{fig:spdupvsF1} for all the classifiers. Note that not in all the cases speedup increases with F1-score. This is due to the effect of the type of misclassification on the overall speedup. For example, if a slow instruction is classified into a faster class, the result at the output of the execution unit at the end of the fast clock period is incorrect. Thus, a  four-clock-cycle penalty is incurred to re-execute the slow instruction, compensating for the combined latency of the re-executed IF, ID, ML, and EX stages. Alternatively, if a fast instruction is misclassified into a slow class, the execution still results in correct answer albeit the potential loss in performance gain. In addition, if a fast instruction is classified into a nominal-delay class (for example, in the case with three delay classes), the overall performance of the system is still increased (but not maximized) as compared with the execution in the slowest delay class (as designed for the worst-case clock period).
To understand the significance of speed and overhead in the overall performance of individual ML classifiers, SPH metric is considered. The SPH results (as determined based on Tables \ref{table:NN}-\ref{table:RF}) are shown in Fig. \ref{fig:hwc_spd} for NN, SVM, and RF classifiers in two-, three-, and four-class configurations. Based on these results, RF exhibits the best tradeoff between the hardware cost and speedup, as well as the lowest design complexity and hardware overheads. RF classifier is, therefore, preferred in this work as a demonstration vehicle of the proposed framework.

\section{Implementation} \label{Implementation}
The proposed framework is implemented with RF model within TigerMIPS and evaluated  based on LegUp benchmarks. The details of the implementation are described in this section.

\subsection{Unified Platform} \label{Unified platform}

A holistic platform is developed based on the proposed system design methodology, as illustrated in Fig. \ref{fig:sysFlow}. The framework is unified within a shell programming platform supported with several peripheral programs developed in C++ and Python. The synthesis steps, as described in Fig. \ref{fig:sysFlow}, are sequentially executed from {\it Start} to {\it Finish}. 

During the first phase, Synopsys Design Compiler is called with the high-level HDL model of the baseline processor.  The profiler triggers are added to the system and GLS is performed in Modelsim. 

The second phase is triggered upon the completion of the instruction profiling. An external parser program is called to transform the instruction profiles into the ML feature data structure and eliminate outliers. The model is trained to classify propagation delays into user-defined number of classes based on a user-specified learning algorithm and delay boundaries. The ML accuracy and estimated speedup are evaluated upon the training completion. If the design requirements are met, the ML software model is transformed into the high-level HDL code. Otherwise, ML model is retrained with new parameters. 

Upon training completion, the HDL code of the ML model is instantiated within the original HDL model of the  baseline processor. Finally, the procedure in Phase 1 is repeated in Phase 3 with the modified processor model, and the overall system performance and overheads are evaluated.

\subsection{Baseline Processor} \label{Baseline processor}
The proposed framework is demonstrated on TigerMIPS. In addition to the basic MIPS units, such as Instruction Fetch (IF), Instruction Decode (ID), Execute (Exe), Memory access (Mem), and Write-back (WB), TigerMIPS comprises advanced units, such as,  forwarding unit, branch handling unit, stall logic, and instruction and data caches, which are common in modern pipeline processors.

\begin{figure*}[t]
  \includegraphics[width=\linewidth]{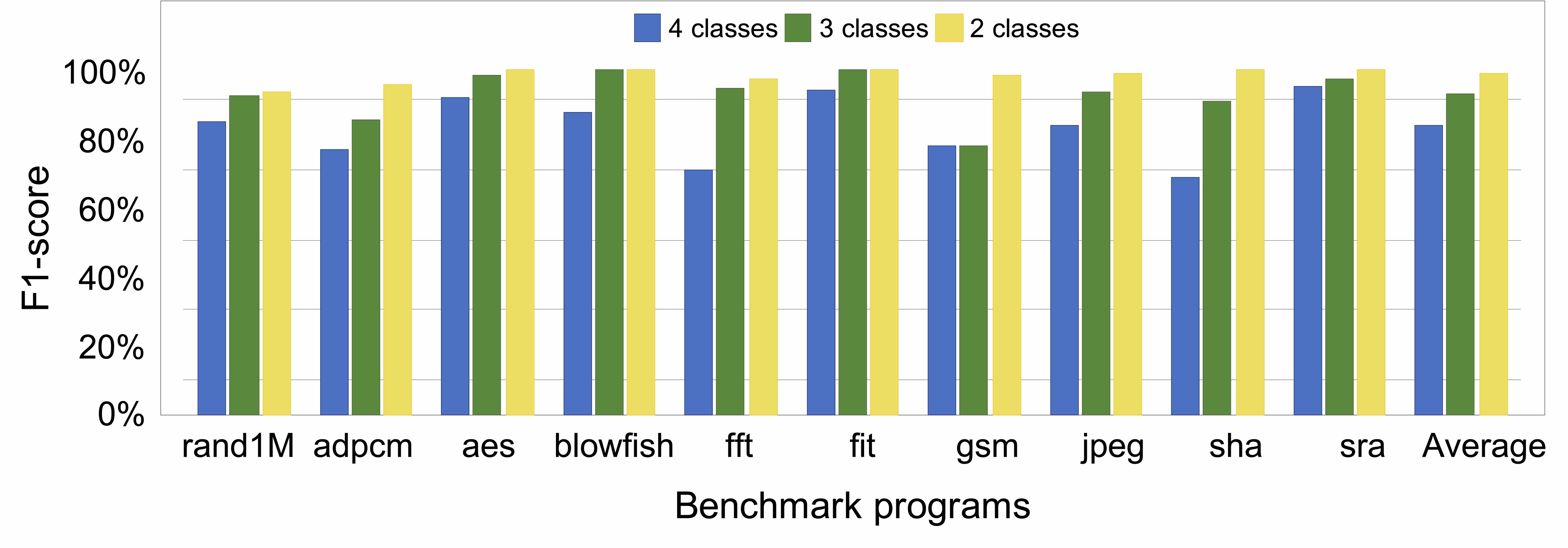}
  \caption{Inference RF classification based on the LegUp benchmark suite with two, three, and four classes. }
  \label{fig:accuracy}
\end{figure*}

\begin{figure*}[!t]
  \includegraphics[width=\linewidth]{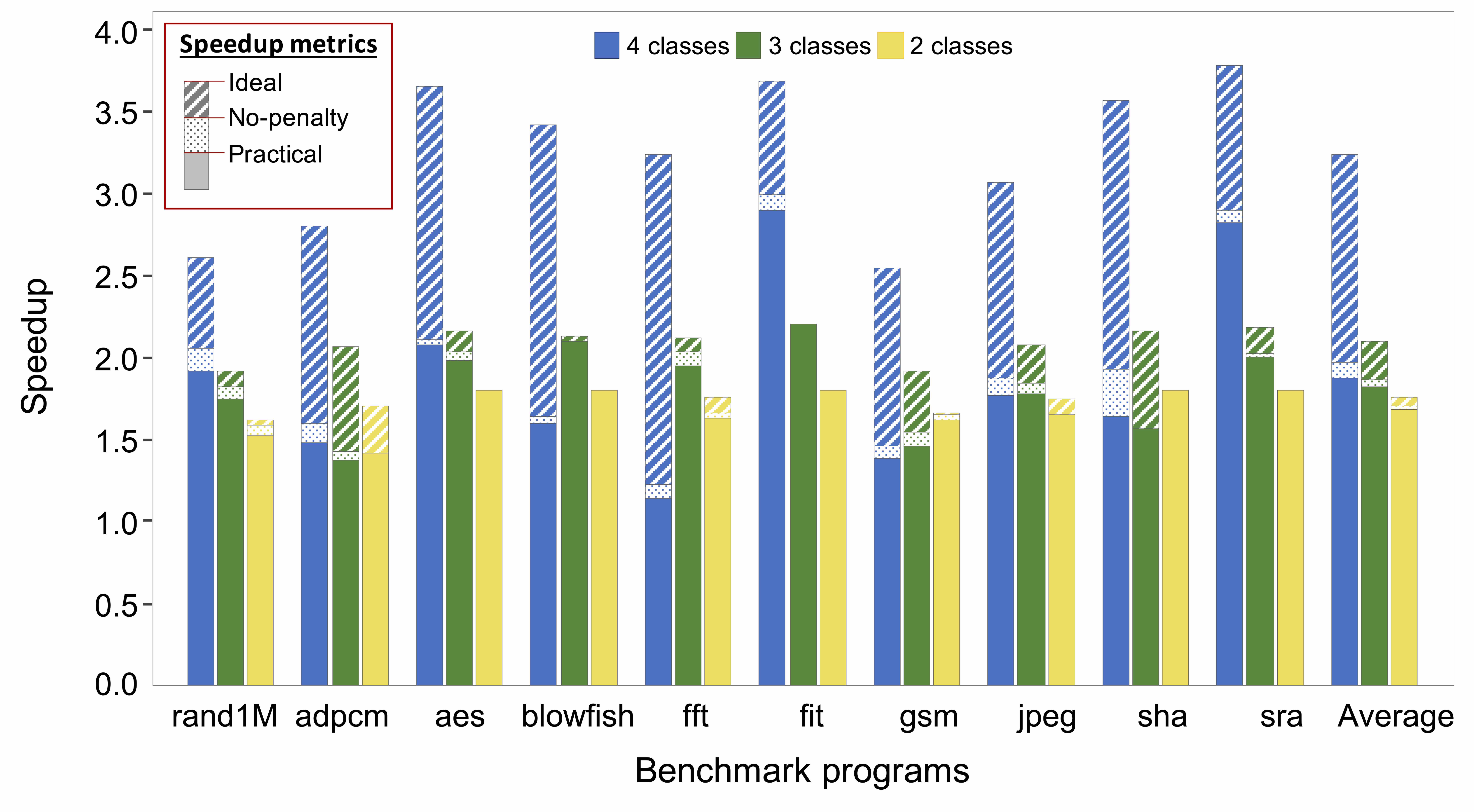}
  \caption{Experimental speedup with the proposed ML framework with two, three, and four delay classes. Practical, no-penalty, and ideal speedups are presented for each benchmark and class. The practical speedup considers the experimental classification accuracy and delay overheads due to misclassification of a slow instruction into a fast class. The no-penalty speedup considers the experimental accuracy, but disregards the idle time due to misclassification of a fast instruction into a slow class. Finally, the ideal speedup is the theoretical maximum with 100\% classification accuracy.}
  \label{fig:speed-up}
\end{figure*}

\subsection{Synthesis and Profiling} \label{Synthesis and profiling}
The baseline model is synthesized in 45 nm NanGate CMOS technology node with Synopsys Design Compiler. Upon completion of the synthesis, triggers are implemented in Verilog HDL, enabling data and timestamp sampling at the input and output of the execution unit within the MIPS pipeline. The profiling is performed based on GLS with Modelsim simulator.

\subsection{Integration, Verification and Evaluation} \label{Integration, Verification and Evaluation}
The trained ML model is first validated in Python. The HDL code of the validated ML model is integrated into the baseline processor.  Finally, the modified processor is synthesized and its functionality is verified through GLS. The post PAR reports are utilized to evaluate the modified system with respect to specified design constraints.

\section{Experimental Results} \label{Experimental Results}

To demonstrate the framework, LegUp high-level synthesis benchmark suite coupled with LLMVM compiler toolchain \cite{llvm} is utilized for profiling and verification during GLS. The trained RF model is tested with nine standard benchmark programs available within the LegUp benchmark suite and an additional synthetically generated benchmark with one million random instructions. The F1-score is shown in Fig. \ref{fig:accuracy} for two, three, and four ML delay classes, yielding above 95\% F1-score for majority of the programs with two delay classes. Resultant speedup for the individual benchmarks is shown in Fig. \ref{fig:speed-up}, including the practical speedup (with the misclassification penalty), no-penalty speedup (without the misclassification penalty), and ideal speedup (with 100\% classification accuracy). The energy overhead due to the additional ML hardware and classification errors is listed in Table \ref{table:energy}. To account for delay overheads due to the misclassification of a slow instruction into a higher performance class, a re-execution penalty of four clock cycles (compensating for IF, ID, ML, and EX stages) is considered within the performance results, as reported in Fig. \ref{fig:speed-up}. The no-penalty speedup is also presented in Fig. \ref{fig:speed-up}, visualizing the penalty due to the misclassification of a fast instruction into a slow class. Note that the overall speedup with four-class configuration is higher than the speedup with two-class configuration, albeit the higher classification accuracy with two delay classes. Alternatively, higher misclassification rate with four delay classes yields higher re-execution energy consumption, as listed in Table \ref{table:energy}. Also, note that a negative energy overhead indicates a reduction in the overall energy consumption ({\it i.e.,} power-delay product).

Performance comparison between the proposed method and state-of-the-art (ML and non-ML) DVFS approaches is listed in Table \ref{table:comparison}. For example, both the proposed framework and the approach in \cite{Hashemi16} consider binary classification with two execution delay classes. The proposed method exhibits 3.5 times higher speedup gain and 33\% energy savings as compared with 3\% energy overhead, as reported in \cite{Hashemi16}. As compared with the adaptive approach in \cite{Jia19_ISSCC}, the proposed method exhibits up to 4.9  times increase in performance gain with 50\% less energy savings. Alternatively, a 3.85 times higher performance gain is demonstrated as compared to \cite{Jia19_ISSCC} with similar energy savings.

\begin{figure}[t]
  \includegraphics[width=\linewidth]{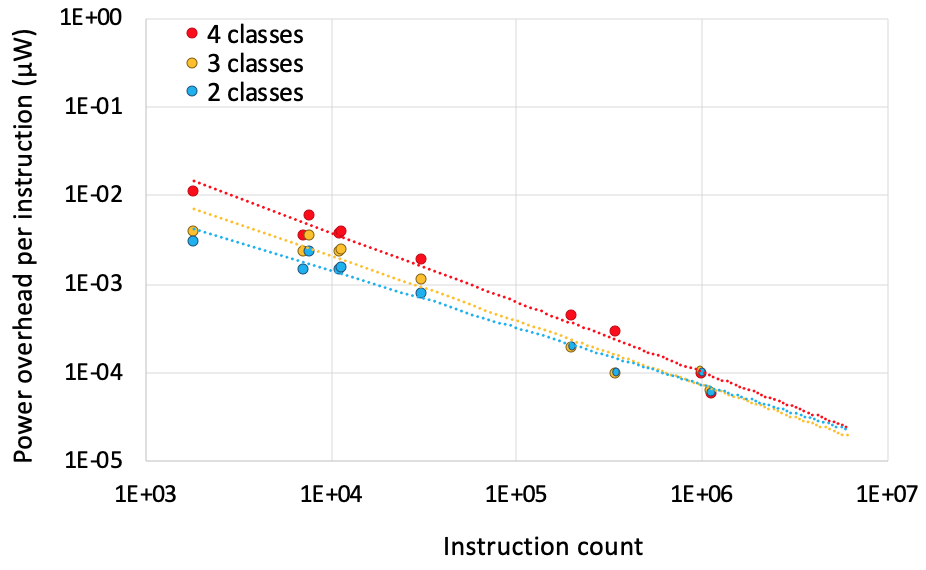}
  \caption{Power overhead per instruction for 2, 3, and 4 delay class configuration based on the benchmarks in Table \ref{table:energy}. }
  \label{fig:power_per_inst}
\end{figure}

Power overhead per instruction for two-, three-, and four-delay class configurations are also determined for the programs in the LegUp benchmark suite. The average power overhead  (due to the additional ML stage and re-execution of misclassified instructions) is shown in Fig. \ref{fig:power_per_inst}. The average power is linearly reduced with the increasing number of program instructions, exhibiting an overhead of less than 0.02 microwatts in practical applications with more than one million instructions. Furthermore, the additional average power consumption rapidly converges for various number of classes, as shown in Fig. 7. Thus, when optimizing the number of delay classes in processors with large workload, power overhead is a secondary factor. Finally, the steeper decrease in the power oberhead with the four-class configuration supports the previous assertion regarding the gain-overhead tradeoff with finer granularity of delay classes: as the number of instructions increases, the higher accuracy with four-class configuration mitigates the adverse effects of misclassifications on the overall system frequency.


\begin{table}[t]
\caption{Experimental power and energy overhead of the proposed ML method. }
\resizebox{\columnwidth}{!}{
\begin{tabular}{| | c ||c  | c | c | c ||}
\hline
    \bfseries    & 
    \multicolumn{4}{|c||}{ \bfseries 4 classes } \\ 
        \cline{2-5} 
     \bfseries  Benchmark   & Practical  & Power     & Energy   	&Instruction \\
    					& speedup & overhead & overhead   &count

    \begin{filecontents*}{result_v2_1.csv}
a,b,c,d,e,f,g,,h,,i,j,k,l,m,n,o,,p,,q,r,s,t,u,v,w,x,y,z,
rand1M,0.849,1.923,2.071,2.633,0.147,0.562,1.385,38.5\%,0.720,-27.99\%,0.923,1.765,1.837,1.933,0.072,0.096,1.233,23.3\%,0.698,-30.151\%,0.937,1.530,1.602,1.631,0.071,0.030,14.29\%,-25.323\%,1000000,rand1M
adpcm,0.776,1.497,1.618,2.819,0.121,1.201,1.565,56.49\%,1.045,4.55\%,0.867,1.389,1.442,2.079,0.053,0.637,1.337,33.69\%,0.962,-3.768\%,0.957,1.418,1.422,1.715,0.005,0.293,22.4\%,-13.654\%,30197,adpcm
aes,0.919,2.087,2.114,3.665,0.027,1.551,1.451,45.14\%,0.695,-30.46\%,0.978,1.987,2.046,2.186,0.059,0.140,1.271,27.06\%,0.639,-36.051\%,1.000,1.818,1.818,1.818,0.000,0.000,17.1\%,-35.595\%,11223,aes
blowfish,0.882,1.633,1.649,3.435,0.015,1.787,1.650,65.01\%,1.010,1.02\%,0.997,2.125,2.125,2.147,0.000,0.022,1.393,39.27\%,0.655,-34.456\%,1.000,1.818,1.818,1.818,0.000,0.000,27.23\%,-30.024\%,199759,blowfish
fft,0.713,1.165,1.226,3.256,0.061,2.030,1.425,42.45\%,1.223,22.28\%,0.942,1.957,2.039,2.135,0.083,0.096,1.255,25.47\%,0.641,-35.872\%,0.973,1.646,1.669,1.776,0.023,0.106,16.14\%,-29.45\%,11001,fft
fir,0.943,2.908,3.006,3.708,0.097,0.702,1.259,25.94\%,0.433,-56.7\%,1.000,2.222,2.222,2.222,0.000,0.000,1.161,16.14\%,0.523,-47.727\%,1.000,1.818,1.818,1.818,0.000,0.000,10.5\%,-39.225\%,7024,fir
gsm,0.782,1.382,1.473,2.562,0.090,1.089,1.458,45.77\%,1.055,5.45\%,0.782,1.465,1.553,1.911,0.088,0.358,1.279,27.87\%,0.873,-12.729\%,0.975,1.635,1.667,1.673,0.031,0.006,18.34\%,-27.642\%,7671,gsm
jpeg,0.840,1.792,1.891,3.078,0.099,1.187,1.550,55.04\%,0.865,-13.49\%,0.937,1.786,1.843,2.095,0.058,0.251,1.326,32.62\%,0.743,-25.74\%,0.987,1.665,1.670,1.766,0.005,0.096,22\%,-26.727\%,1133161,jpeg
sha,0.694,1.657,1.935,3.585,0.278,1.650,1.621,62.07\%,0.978,-2.22\%,0.907,1.578,1.579,2.178,0.001,0.599,1.379,37.93\%,0.874,-12.566\%,1.000,1.818,1.818,1.818,0.000,0.000,27.23\%,-30.024\%,345576,sha
sra,0.953,2.840,2.907,3.793,0.067,0.885,1.206,20.62\%,0.425,-57.53\%,0.966,2.013,2.037,2.202,0.024,0.166,1.072,7.18\%,0.533,-46.745\%,1.000,1.818,1.818,1.818,0.000,0.000,5.5\%,-41.975\%,1775,sra
\end{filecontents*}
\csvreader[head to column names]{result_v2_1.csv}{}
    {\\\hline \a &  \c & \h & \i & \z }
        \\   \Cline{0.65pt}{1-5}
   \multicolumn{1}{||c||}{ Average }		        &  1.889	&	45.7\%	&	-15.51\%	&	274738.7  	\\
   \cline{1-5}
   \multicolumn{1}{||c||}{ Positive}		 	&  		&			&			&		   		\\
   \multicolumn{1}{||c||}{ standard }	  		&   0.35	&	5.81\%	&	8.70\%	&	374831.68   	\\
   \multicolumn{1}{||c||}{ deviation($\sigma^+$)} 	&  		&			&			&		    		\\
   \cline{1-5}
   \multicolumn{1}{||c||}{ Negative}		 	&  		&			&			&		    		\\
   \multicolumn{1}{||c||}{ standard }	  		&   0.17	&	6.58\%	&	15.50\%	&	92682.62    	\\
   \multicolumn{1}{||c||}{ deviation($\sigma^-$)} 	&  		&			&			&		   
 \\\hline   \hline

   \bfseries    & 
    \multicolumn{4}{|c||}{ \bfseries 3 classes } \\ 
        \cline{2-5} 
     \bfseries  Benchmark   & Practical  & Power     & Energy   	&Instruction \\
    					& speedup & overhead & overhead   &count

    \csvreader[head to column names]{result_v2_1.csv}{}
    {\\\hline \a & \k & \p & \q & \z }
        \\   \Cline{0.65pt}{1-5}
   \multicolumn{1}{||c||}{ Average }		        &  1.829	&	27.05\%	&	-28.58\%	&	274738.7  	\\
   \cline{1-5}
   \multicolumn{1}{||c||}{ Positive}		 	&  		&			&			&		  		\\
   \multicolumn{1}{||c||}{ standard }	  		&   0.11	&	3.09\%	&	8.41\%	&	374831.68   	\\
   \multicolumn{1}{||c||}{ deviation($\sigma^+$)} 	&  		&			&			&		    		\\
   \cline{1-5}
   \multicolumn{1}{||c||}{ Negative}		 	&  		&			&			&		    		\\
   \multicolumn{1}{||c||}{ standard }	  		&   0.13	&	5.76\%	&	4.84\%	&	92682.62    	\\
   \multicolumn{1}{||c||}{ deviation($\sigma^-$)} 	&  		&			&			&		   
 \\\hline   \hline

   \bfseries    & 
    \multicolumn{4}{|c||}{ \bfseries 2 classes } \\ 
        \cline{2-5} 
     \bfseries  Benchmark   & Practical  & Power     & Energy   	&Instruction \\
    					& speedup & overhead & overhead   &count

    \csvreader[head to column names]{result_v2_1.csv}{}
    {\\\hline \a &  \s & \x & \y & \z}
        \\   \Cline{0.65pt}{1-5}
   \multicolumn{1}{||c||}{ Average }		        &  1.699	&	18.073\%	&	-29.964\%	&	274738.7		\\
   \cline{1-5}
   \multicolumn{1}{||c||}{ Positive}		 	&  		&			&			&		   		\\
   \multicolumn{1}{||c||}{ standard }	  		&   0.05	&	2.84\%	&	3.49\%	&	374831.68	\\
   \multicolumn{1}{||c||}{ deviation($\sigma^+$)} 	&  		&			&			&		    		\\
   \cline{1-5}
   \multicolumn{1}{||c||}{ Negative}		 	&  		&			&			&		    		\\
   \multicolumn{1}{||c||}{ standard }	  		&   0.07	&	3.06\%	&	3.24\%	&	92682.62		\\
   \multicolumn{1}{||c||}{ deviation($\sigma^-$)} 	&  		&			&			&		   
    \\\hline

\end{tabular}
}
\label{table:energy}
\end{table}

\section{Conclusions and Future Work} \label{Discussion}

The proposed unified framework facilitates efficient utilization of the time and hardware recourses in the system. In addition, this approach enables the design of ML pipeline stages, while satisfying design constraints, as shown in Fig. \ref{fig:sysFlow}. Finally, classification of instructions into delay intervals in real time alleviates the path propagation variances imposed by PVT variations and system aging. To enhance the performance gain, the proposed approach should be preferred with those applications and systems characterized by considerable variations in the propagation delay of the individual instructions. 

This method is  practical with pipelined, MIPS-like processors, in which the overall delay is dominated by the delay of the execution stage. Although, the proposed method is explored in this work with a single core system, further increases in energy efficiency and the overall system performance are expected if the approach is adjusted for modern architecture processors with out-of-order execution and multicore processors with multiple frequency domains. To exploit the positive impact of out-of-order execution and multicore systems on performance and energy efficiency in commercial class processors, the following methodologies  should be considered.

\begin{table}[t]
\small\addtolength{\tabcolsep}{-1pt}
\caption{Comparison between the proposed method and existing state-of-the-art methods.}
\resizebox{\columnwidth}{!}{
\begin{tabular}{||c c| c | c | c | c |  c | c ||} 
 \hline
  Algorithm				&& Performance 					& Energy 				& ML  	\\ [0.5ex] 
  		    				&& 				 gain				& 		 overhead		&  based 	\\ [0.5ex] 
 \hline\hline
 
 \multicolumn{2}{||c|}{SLoT \cite{rahimi17_slot}		}& 23\%  						& N/A 				&  Yes 	  \\

\hline		
 \multicolumn{2}{||c|}{Early Prediction \cite{Hashemi16}}& 20\%					   	&  3\% 				& No  \\

\hline		
 \multicolumn{2}{||c|}{ Clim \cite{rahimi17_clim}		}& 24\%   						&  N/A				&  Yes\\
		
\hline		
 \multicolumn{2}{||c|}{ SLBM \cite{rahimi15}		}& 15\%   						&  N/A				&  Yes\\
\hline		
 \multicolumn{2}{||c|}{ Adaptive Clock 			 }& 18.2\%  	 			 	&  -30.4\%				 &  No\\	
 \multicolumn{2}{||c|}{Management\cite{Jia19_ISSCC}}& 	   						&  	  				 &  \\

 \hline

\multicolumn{1}{||c|}{  }		  		&  2 classes      &  70\% 				&	-30\%	& \multicolumn{1}{c|}{  } \\
\cline{2-4}						
\multicolumn{1}{||c|}{ This work }  		&  3 classes   	&   83\%  				&	-28.6\%		& \multicolumn{1}{c|}{ Yes } \\
\cline{2-4}
\multicolumn{1}{||c|}{  }		   		&  4 classes   	&   89\%  				&	-15.5\%		& \multicolumn{1}{c|}{  } \\
		
 \hline

\end{tabular}
}
\label{table:comparison}
\end{table}

\subsection{Single-Core, Single-Clock Delay-Based Out-of-Order Execution} \label{one}
To support out-of-order execution, instructions within a delay class should be bundled into a delay-class specific reservation station (RS). Instructions stored in an RS are individually executed at a constant frequency until the RS is emptied or a dependency is determined, preventing further execution of instructions in the RS. Such bundling of instructions reduces the number of clock signal transitions among various frequencies, increasing the performance and power efficiency of the system.
\subsection {Single-Core, Multi-Clock Delay-Based Out-of-Order Execution} \label{two}
As previously, to support out-of-order execution, instructions should be bundled based on the delay classes and stored within the matching RS's. To support multi-clock execution, the ALUs and FPUs within the execution unit should be operated at different clock frequencies, as determined by the granularity of the delay classes. Intuitively, the parallelization of execution from different delay classes with this approach decreases the number of clock adjustments, increasing the system performance and energy efficiency.
\subsection{Multi-Core, Multi-Clock Delay-Based Out-of-Order Execution} 
To leverage the advantages provided by processing with multiple clock domains in multicore systems, bundled instructions within the individual clock domains (as defined in subsection \ref{one}) should be shared among all the system clock domains, mitigating the additional cost of multiple clocking (as described in subsection \ref{two}). To enable the sharing of bundles, efficient bundle scheduling and low overhead communication channels are required. While the number of clock adjustments is expected to further reduce with this approach, additional overheads due to intelligent communication of bundles among the cores should be considered. Alternatively, by partially or fully replacing the traditional DFS, DVFS, and thread scheduling mechanisms, additional savings are expected with the proposed approach. Finally, the proposed method can be adjusted in a similar manner to classify instruction propagation delay of various pipeline stages.

Existing approaches are focused on offline speculations, statistical models, per-task (workload-based) frequency scaling, and prediction of timing errors at an operating point of a system. Alternatively, the proposed method demonstrates the benefits of fine-grain, instruction-level frequency adjustment, simultaneously utilizing most of the clock period slack and mitigating  the adverse effects of PVT variations and aging.

\section{Summary} \label{Conclusion}
In this work, an additional ML pipeline stage is proposed for increasing the overall system performance by enhancing the temporal resource utilization. This additional stage is designed to classify instructions into propagation delay classes. The system clock frequency is adaptively adjusted based on the individual delay class predictions. Pipelining is exploited to mitigate the effect of the ML stage latency on the overall system performance. Practical ML features are extracted based on current instruction and computation history. ML hardware and misclassification power and delay overheads are considered within the reported results. TigerMIPS is utilized as the baseline processor. The processor is enhanced with the ML predictor and simulated with the  LegUp benchmark suite. Based on the experimental results, up to 89\% performance gain is achieved with four delay classes with 15.5\% energy saving. Alternatively, the reduction of 30\% in energy consumption with 70\% performance gain is demonstrated with two delay classes. A unified shell programing platform with peripheral programs is designed to provide a systematic design flow for ML driven pipelined processors.

\begin{IEEEbiography}[{\includegraphics[width=1in,height=1.25in,clip,keepaspectratio]{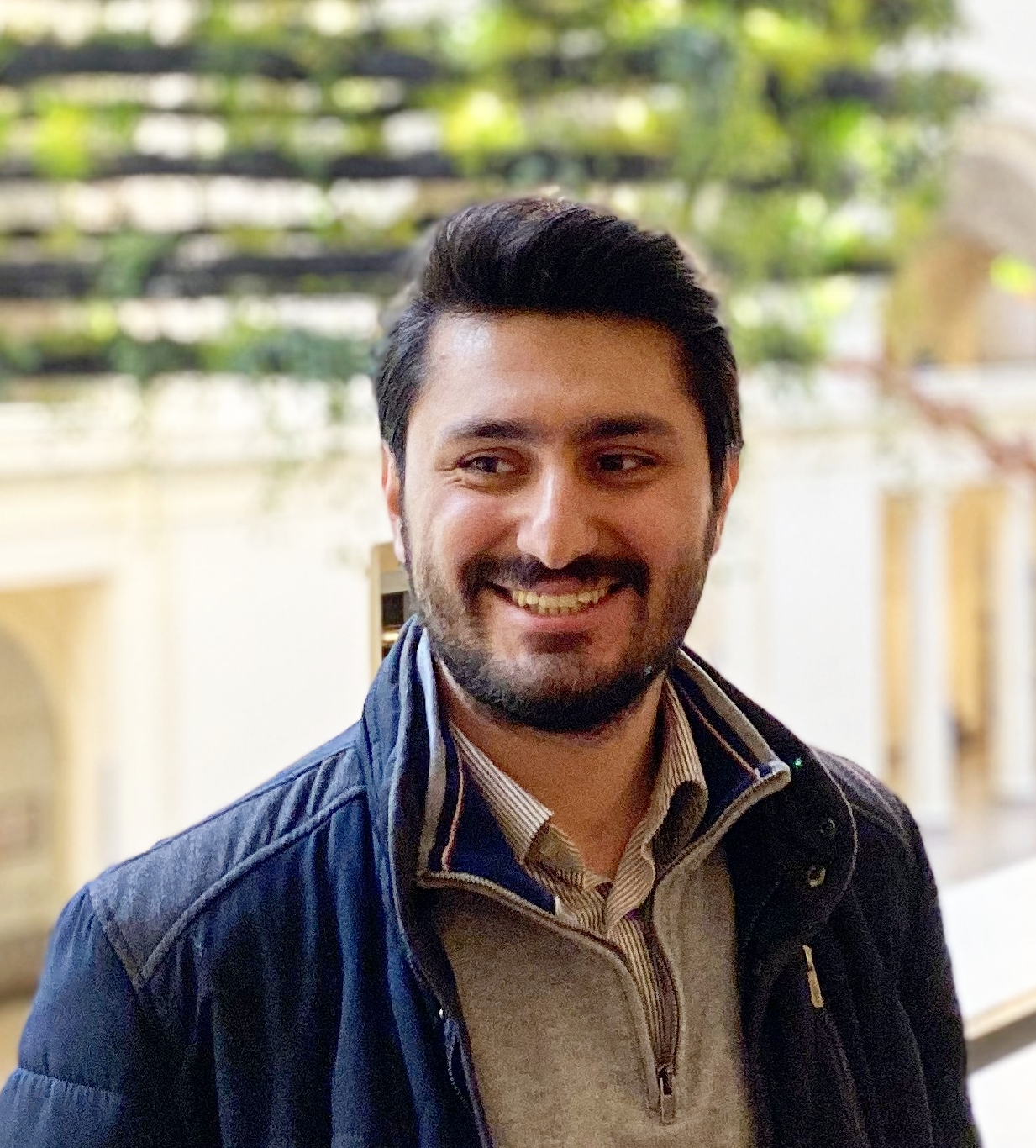}}]{Arash Fouman Ajirlou}

(S'17) received the Bachelor of Science degree in computer engineering from University of Tehran, Tehran, Iran, in 2017. He started the PhD program with Department of Electrical and Computer Engineering at the University of Illinois at Chicago, in 2018. He was a research assistant in the school of Electrical and Computer Engineering at University of Tehran between 2015 and late 2017. From 2017 to late 2018, he served as the secretary of the Electrical and Computer Engineering committee in Alumni Association of Faculty of Engineering, University of Tehran. In 2018, prior to starting his PhD in computer engineering at University of Illinois at Chicago, he was a digital designer in the engineering department of Ofogh Tajrobe Moj company, Tehran, Iran.

His primary interests are embedded systems and high-performance/low-power computing systems, with an emphasis on machine learning and self governing systems. His current focus is on utilizing machine learning methodologies to enhance processor performance and energy consumption.
\end{IEEEbiography}

\begin{IEEEbiography}[{\includegraphics[width=1in,height=1.25in,clip,keepaspectratio]{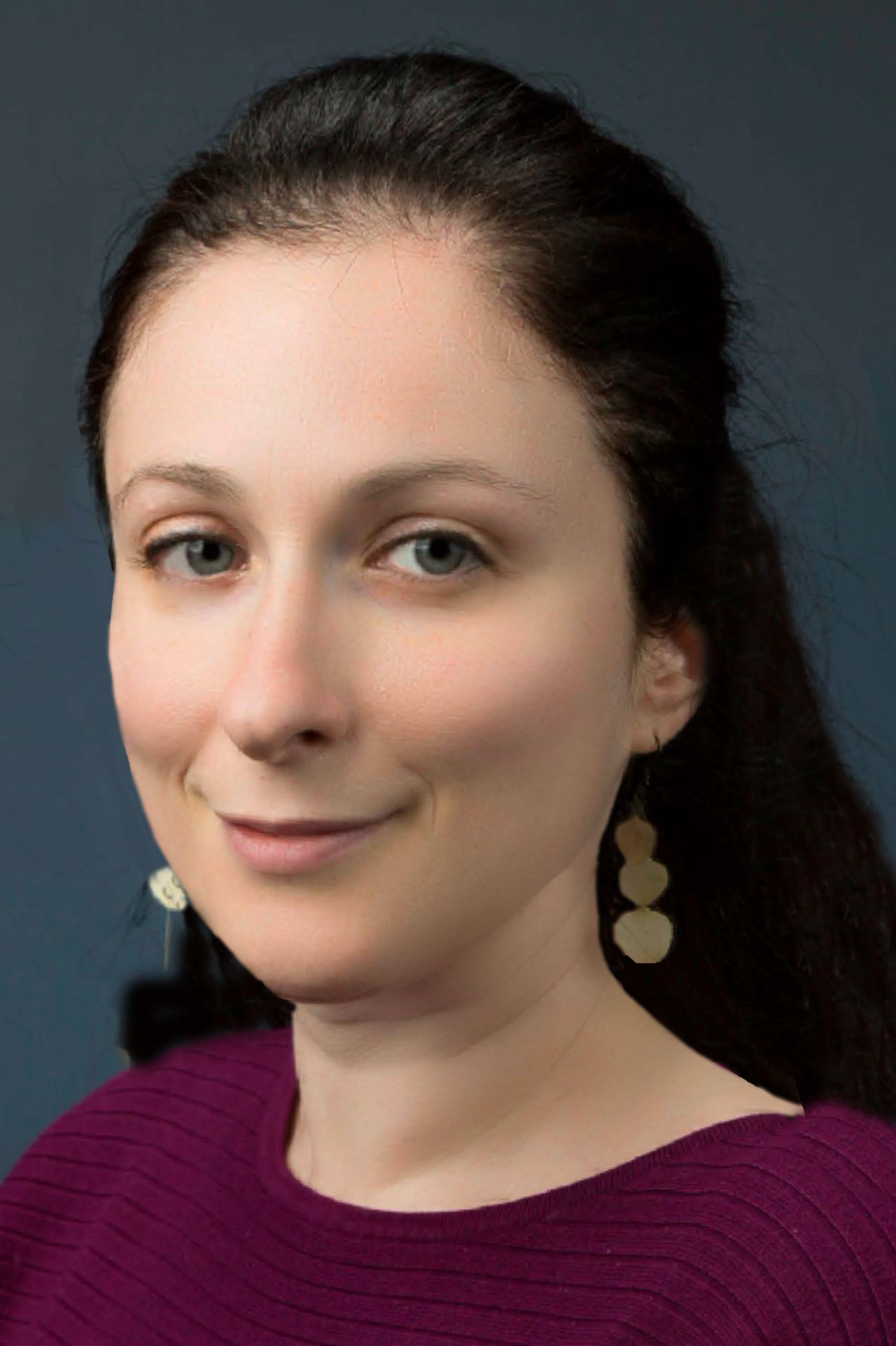}}]%
{Dr. Inna Partin-Vaisband}

(S'12--M'15) received the Bachelor of Science degree in computer engineering and the Master of Science degree in electrical engineering from the Technion-Israel Institute of Technology, Haifa, Israel, in, respectively, 2006 and 2009, and the Ph.D. degree in electrical engineering from the University of Rochester, Rochester, NY in 2015. She is currently an Assistant Professor with the Department of Electrical and Computer Engineering at the University of Illinois at Chicago. 

Between 2003 and 2009, she held a variety of software and hardware R\&D positions at Tower Semiconductor Ltd., GConnect Ltd., and IBM Ltd., all in Israel. Her primary interests lay in the area of high performance integrated circuits and VLSI system design. Her research is currently focused on innovation in the areas of AI hardware and hardware security. Yet another primary focus is on distributed power delivery and locally intelligent power management that facilitates performance scalability in heterogeneous ultra-large scale integrated systems. Special emphasis is placed on developing robust frameworks across levels of design abstraction for complex heterogeneous integrated systems. Dr. P.-Vaisband is an Associate Editor of the Microelectronics Journal and has served on the Technical Program and Organization Committees of various conferences.
\end{IEEEbiography}


\begin{thebibliography}{10}

\bibitem{Fields02}
Fields B, Bodík R, Hill MD. Slack: Maximizing performance under technological constraints. InProceedings 29th Annual International Symposium on Computer Architecture 2002 May 25 (pp. 47-58). IEEE.

\bibitem{Zyuban04}
Zyuban V, Brooks D, Srinivasan V, Gschwind M, Bose P, Strenski PN, Emma PG. Integrated analysis of power and performance for pipelined microprocessors. IEEE Transactions on Computers. 2004 Jun 21;53(8):1004-16.


\bibitem{kumar03}
Kumar R, Farkas KI, Jouppi NP, Ranganathan P, Tullsen DM. Single-ISA heterogeneous multi-core architectures: The potential for processor power reduction. InProceedings of the 36th annual IEEE/ACM International Symposium on Microarchitecture 2003 Dec 3 (p. 81). IEEE Computer Society.

\bibitem{rahimi17_slot}
Jiao X, Jiang Y, Rahimi A, Gupta RK. Slot: A supervised learning model to predict dynamic timing errors of functional units. InProceedings of the Conference on Design, Automation \& Test in Europe 2017 Mar 27 (pp. 1183-1188). European Design and Automation Association.

\bibitem{Hashemi16}
Hashemi SH, Ajirlou AF, Soltani M, Navabi Z. Early prediction of timing critical instructions in pipeline processor. In2016 15th Biennial Baltic Electronics Conference (BEC) 2016 Oct 3 (pp. 95-98). IEEE.

\bibitem{Moghadas18}
Moghaddasi I, Fouman A, Salehi ME, Kargahi M. Instruction-level NBTI Stress Estimation and its Application in Runtime Aging Prediction for Embedded Processors. IEEE Transactions on Computer-Aided Design of Integrated Circuits and Systems. 2018 Jun 12.

\bibitem{gepner06} 
Gepner P, Kowalik MF. Multi-core processors: New way to achieve high system performance. InInternational Symposium on Parallel Computing in Electrical Engineering (PARELEC'06) 2006 Sep 13 (pp. 9-13). IEEE.



\bibitem{hu04}
Hu Z, Buyuktosunoglu A, Srinivasan V, Zyuban V, Jacobson H, Bose P. Microarchitectural techniques for power gating of execution units. InProceedings of the 2004 international symposium on Low power electronics and design 2004 Aug 9 (pp. 32-37). ACM.

\bibitem{wu04}
Wu Q, Pedram M, Wu X. Clock-gating and its application to low power design of sequential circuits. IEEE Transactions on Circuits and Systems I: Fundamental Theory and Applications. 2000 Mar;47(3):415-20.

\bibitem{Wang19}
Wang S, Ananthanarayanan G, Zeng Y, Goel N, Pathania A, Mitra T. High-throughput cnn inference on embedded arm big. little multi-core processors. IEEE Transactions on Computer-Aided Design of Integrated Circuits and Systems. 2019 Sep 30.

\bibitem{Rapp19}
Rapp M, Sagi M, Pathania A, Herkersdorf A, Henkel J. Power-and Cache-Aware Task Mapping with Dynamic Power Budgeting for Many-Cores. IEEE Transactions on Computers. 2019 Aug 20;69(1):1-3.

\bibitem{Isci06}
Isci C, Buyuktosunoglu A, Buyuktosunoglu A, Cher CY, Bose P, Martonosi M. An analysis of efficient multi-core global power management policies: Maximizing performance for a given power budget. InProceedings of the 39th annual IEEE/ACM international symposium on microarchitecture 2006 Dec 9 (pp. 347-358). IEEE Computer Society.

\bibitem{legup}
Canis A, Choi J, Aldham M, Zhang V, Kammoona A, Anderson JH, Brown S, Czajkowski T. LegUp: high-level synthesis for FPGA-based processor/accelerator systems. InProceedings of the 19th ACM/SIGDA international symposium on Field programmable gate arrays 2011 Feb 27 (pp. 33-36). ACM.


\bibitem{rahimi17_clim}
Jiao X, Rahimi A, Jiang Y, Wang J, Fatemi H, De Gyvez JP, Gupta RK. Clim: A cross-level workload-aware timing error prediction model for functional units. IEEE Transactions on Computers. 2017 Dec 14;67(6):771-83.

\bibitem{FATE18}
Zhang JJ, Garg S. FATE: fast and accurate timing error prediction framework for low power DNN accelerator design. In2018 IEEE/ACM International Conference on Computer-Aided Design (ICCAD) 2018 Nov 5 (pp. 1-8). IEEE.

\bibitem{rahimi15}
Jiao X, Rahimi A, Narayanaswamy B, Fatemi H, de Gyvez JP, Gupta RK. Supervised learning based model for predicting variability-induced timing errors. In2015 IEEE 13th International New Circuits and Systems Conference (NEWCAS) 2015 Jun 7 (pp. 1-4). IEEE.

\bibitem{BandiTS17}
Zhang JJ, Garg S. BandiTS: dynamic timing speculation using multi-armed bandit based optimization. InDesign, Automation \& Test in Europe Conference \& Exhibition (DATE), 2017 2017 Mar 27 (pp. 922-925). IEEE.

\bibitem{MAB}
Whittle P. Multi-armed bandits and the Gittins index. Journal of the Royal Statistical Society: Series B (Methodological). 1980 Jan;42(2):143-9.

\bibitem{salamat19}
Khaleghi B, Salamat S, Imani M, Rosing T. FPGA Energy Efficiency by Leveraging Thermal Margin. arXiv preprint arXiv:1911.07187. 2019 Nov 17.

\bibitem{assare19}
Assare O, Gupta R. Accurate Estimation of Program Error Rate for Timing-Speculative Processors. InProceedings of the 56th Annual Design Automation Conference 2019 2019 Jun 2 (p. 180). ACM.

\bibitem{Kruijf10}
De Kruijf M, Nomura S, Sankaralingam K. A unified model for timing speculation: Evaluating the impact of technology scaling, CMOS design style, and fault recovery mechanism. In2010 IEEE/IFIP International Conference on Dependable Systems \& Networks (DSN) 2010 Jun 28 (pp. 487-496). IEEE.


\bibitem{TigerMIPS}
Moore, S. and Chadwick, G., 2011. The Tiger ``MIPS'' processor.

\bibitem{scikit-learn}
Pedregosa F, Varoquaux G, Gramfort A, Michel V, Thirion B, Grisel O, Blondel M, Prettenhofer P, Weiss R, Dubourg V, Vanderplas J. Scikit-learn: Machine learning in Python. Journal of machine learning research. 2011;12(Oct):2825-30.

\bibitem{Jia19_ISSCC}
Jia T, Joseph R, Gu J. 19.4 An Adaptive Clock Management Scheme Exploiting Instruction-Based Dynamic Timing Slack for a General-Purpose Graphics Processor Unit with Deep Pipeline and Out-of-Order Execution. In2019 IEEE International Solid-State Circuits Conference-(ISSCC) 2019 Feb 17 (pp. 318-320). IEEE.

\bibitem{James13}
G. James, et al., ``An Introduction to Statistical Learning,'' New York: Springer, Vol. 112, 2013.

\bibitem{Kuhn13}
M. Kuhn and J. Kjell, ``Applied Predictive Modeling,'' New York: Springer, Vol. 26, 2013.

\bibitem{Kohavi}
R. Kohavi, ``A Study of Cross-Validation and Bootstrap for Accuracy Estimation and Model Selection,'' Proc. of the International Joint Conference on Artificial Intelligence, Vol. 14, No. 2, pp. 1137-114, 1995.

\bibitem{Forman}
G. Forman and S. Scholtz, ``Apples-to-Apples in Cross-Validation Studies: Pitfalls in Classifier Performance Measurement.'' ACM SIGKDD Explorations Newsletter, Vol. 12, No. 1, pp. 49-57, 2010.


\bibitem{NN1}
Yue J, Liu R, Sun W, Yuan Z, Wang Z, Tu YN, Chen YJ, Ren A, Wang Y, Chang MF, Li X. 7.5 A 65nm 0.39-to-140.3 TOPS/W 1-to-12b Unified Neural Network Processor Using Block-Circulant-Enabled Transpose-Domain Acceleration with 8.1× Higher TOPS/mm 2 and 6T HBST-TRAM-Based 2D Data-Reuse Architecture. In2019 IEEE International Solid-State Circuits Conference-(ISSCC) 2019 Feb 17 (pp. 138-140). IEEE.

\bibitem{NN2}
Lee J, Lee J, Han D, Lee J, Park G, Yoo HJ. 7.7 lnpu: A 25.3 tflops/w sparse deep-neural-network learning processor with fine-grained mixed precision of fp8-fp16. In2019 IEEE International Solid-State Circuits Conference-(ISSCC) 2019 Feb 17 (pp. 142-144). IEEE.

\bibitem{NN3}
Lee J, Lee J, Han D, Lee J, Park G, Yoo HJ. 7.7 lnpu: A 25.3 tflops/w sparse deep-neural-network learning processor with fine-grained mixed precision of fp8-fp16. In2019 IEEE International Solid-State Circuits Conference-(ISSCC) 2019 Feb 17 (pp. 142-144). IEEE.

\bibitem{gradientDescent}
Ruder S. An overview of gradient descent optimization algorithms. arXiv preprint arXiv:1609.04747. 2016 Sep 15.

\bibitem{bfgs}
Liu DC, Nocedal J. On the limited memory BFGS method for large scale optimization. Mathematical programming. 1989 Aug 1;45(1-3):503-28.

\bibitem{adam}
Kingma DP, Ba J. Adam: A method for stochastic optimization. arXiv preprint arXiv:1412.6980. 2014 Dec 22.

\bibitem{transistorCount}
Asadi P, Navi K. A new low power 32× 32-bit multiplier. World Applied Sciences Journal. 2007;2(4):341-7.

\bibitem{kernelTrick}
Hofmann M. Support vector machines-kernels and the kernel trick. Notes. 2006 Jun 26;26(3).

\bibitem{svmCost}
Mitéran J, Bouillant S, Bourennane E. Classification boundary approximation by using combination of training steps for real-time image segmentation. InInternational Workshop on Machine Learning and Data Mining in Pattern Recognition 2003 Jul 5 (pp. 141-155). Springer, Berlin, Heidelberg.

\bibitem{mohsenin16}
Kulkarni A, Pino Y, Mohsenin T. SVM-based real-time hardware Trojan detection for many-core platform. In2016 17th International Symposium on Quality Electronic Design (ISQED) 2016 Mar 15 (pp. 362-367). IEEE.


\bibitem{RF}
Zhang X, Wang W, Zheng X, Ma Y, Wei Y, Li M, Zhang Y. A Clutter Suppression Method Based on SOM-SMOTE Random Forest. In2019 IEEE Radar Conference (RadarConf) 2019 Apr 22 (pp. 1-4). IEEE.

\bibitem{comparatorCost}
Cheng SW. A high-speed magnitude comparator with small transistor count. In10th IEEE International Conference on Electronics, Circuits and Systems, 2003. ICECS 2003. Proceedings of the 2003 2003 Dec 14 (Vol. 3, pp. 1168-1171). IEEE.


\bibitem{llvm}
Lattner C, Adve V. LLVM: A compilation framework for lifelong program analysis \& transformation. InProceedings of the international symposium on Code generation and optimization: feedback-directed and runtime optimization 2004 Mar 20 (p. 75). IEEE Computer Society.


\bibitem{Agarwal06}
Agarwal K, Sylvester D, Blaauw D. Modeling and analysis of crosstalk noise in coupled RLC interconnects. IEEE Transactions on Computer-Aided Design of Integrated Circuits and Systems. 2006 Apr 24;25(5):892-901.


\end{thebibliography}
\end{document}